\documentclass[aps,prd,eqsecnum,amsmath,amssymb,longbibliography,10pt,notitlepage,twocolumn]{revtex4-1}
\usepackage[utf8]{inputenc}
\usepackage[T1]{fontenc} 
\usepackage[colorlinks=true, linkcolor=blue, citecolor=teal]{hyperref}
\usepackage{bm}
\usepackage{latexsym,amssymb,amsmath,amsfonts}
\usepackage{graphicx}
\usepackage{physics}
\usepackage{bm}
\usepackage{footnotebackref}
\usepackage{multirow}
\usepackage[usenames]{xcolor}
\usepackage{soul}
\usepackage{makecell}
\usepackage{graphicx}
\usepackage{amssymb}  
\usepackage{amsmath}
\usepackage{rotating}
\usepackage{color}
\usepackage{ifthen}
\usepackage{slashed}
\usepackage{xspace}
\usepackage{tabularx}
\usepackage{float}
\usepackage[normalem]{ulem}
\def\be{\begin{equation}}
\def\ee{\end{equation}}
\def\ba{\begin{aligned}}
\def\ea{\end{aligned}}
\def\bi{\begin{itemize}}
\def\ei{\end{itemize}}
\def\ben{\begin{enumerate}}
\def\een{\end{enumerate}}

\def\d{{\rm d}}

\newcommand{\hatom}{\hat{\Omega}}

\newcommand{\lmax}{\ell_{\rm max}}
\newcommand{\lgwb}{\ell_{\rm max}^{\rm gwb}}
\newcommand{\lrec}{\ell_{\rm max}^{\rm rec}}
\newcommand{\lres}{\ell_{\rm max}^{\rm res}}

\newcommand{\lNpair}{\ell_{\rm max}^{N_{\rm pair}}}

\usepackage{booktabs}

\begin{document}

\title{Equivalence of mapping gravitational wave background anisotropy across bases at equal information content}

\author{Deepali Agarwal}
\email{deepali.agarwal@utrgv.edu}
\affiliation{Department of Physics and Astronomy, University of Texas Rio Grande Valley, One West University Boulevard, Brownsville, Texas 78520, USA}

\begin{abstract}
Mapping anisotropies in the gravitational-wave background (GWB) requires choosing a basis to represent the sky intensity, such as pixel and spherical harmonic bases. Each introduces a truncation---e.g., a number of pixels or a maximum multipole $\lmax$---often set by a naive counting argument that relate the number of measurable modes to the number of independent cross-correlations, $N_{\rm pair}$, in a pulsar timing array (PTA). However, such truncations can lead to reconstruction artifacts, as they do not reflect the true information content of the PTA response. Increasing the value of truncation parameters spans the observable space better but reveals poorly constrained modes, making the inverse problem ill conditioned and requiring regularization. A natural approach is to restrict the reconstruction to a well-measured subspace via principal maps, defined by the dominant eigenmodes of the detector response (or Fisher) matrix. However, these maps are not a fundamental parametrization of the sky but rather, are derived from an underlying representation---such as a pixelization or a spherical harmonic expansion. While their explicit form depends on basis choice, they can span the \textit{same subspace} when the underlying representation is sufficiently complete. Here, we show that reconstructed anisotropy ``maps'' via different bases are equivalent, provided they retain the \textit{same information} content, i.e., span the same principal subspace. As illustrative cases, we consider a toy model for PTA configuration and several GWB anisotropy shapes: point source, an extended source with deterministic anisotropy, and a statistical isotropic background, along with its summary statistic---the angular power spectrum. Although the reconstructions are equivalent, their computational costs can differ. We conclude with brief comments on the construction of principal maps for ground-based interferometers.
\end{abstract}

\maketitle

\section{Introduction}

Recent hints of a gravitational wave background (GWB) signal in pulsar timing array (PTA) observations~\cite{Reardon:2023gzh,EPTA:2023fyk,Miles:2024seg,NANOGrav:2023gor,Xu:2023wog} have motivated the development of improved anisotropy reconstruction methods and a reexamination of systematic effects in existing approaches. A central question is whether the choice of sky basis used to decompose anisotropy affect the reconstruction results~\cite{Konstandin:2025ifn,Mingarelli:2026}.

\begin{table*}[]
\centering
\renewcommand{\arraystretch}{1.5} 
\setlength{\tabcolsep}{10pt}    
\resizebox{\textwidth}{!}{%
\begin{tabular}{c|c|c|c}
\toprule
\textbf{GWB Directional Power} & \multicolumn{3}{c}{$P(\hat{\Omega})$} \\
\midrule
\textbf{Cross-correlations; Pulsar $a$ and $b$} & \multicolumn{3}{c}{$\langle\rho_{ab}\rangle = \int \mathrm{d}^2\hat{\Omega} \, \gamma_{ab}(\hat{\Omega}) \, P(\hat{\Omega})$} \\
\midrule
\textbf{Pairwise Timing Response (PTR)}~\cite{Ali-Haimoud:2020ozu,alihaimoud2021} & \multicolumn{3}{c}{$\gamma_{ab}(\hat{\Omega})$}\\
\midrule
\multirow{1}{*}{\textbf{Basis-independent Maximum Likelihood (ML) Estimation}} 
 & \multicolumn{3}{c}{$X(\hat{\Omega}) = \int \mathrm{d}^2\hat{\Omega}' \, F(\hat{\Omega},\hat{\Omega}') \, \hat{P}(\hat{\Omega}')$} \\
\multirow{2}{*}{Assumption $\rho_{ab}\sim \mathcal{N}(\langle \rho_{ab}\rangle,\sigma_{ab})$}  & \textbf{Fisher kernel} & \multicolumn{2}{c}{$F(\hat{\Omega},\hat{\Omega}') \equiv \sum_{ab} \frac{\gamma_{ab}(\hat{\Omega}) \, \gamma_{ab}(\hat{\Omega}')}{\sigma^2_{ab}}$} \\
 & \textbf{Dirty map} & \multicolumn{2}{c}{$X(\hat{\Omega}) \equiv \sum_{ab} \frac{\gamma_{ab}(\hat{\Omega}) \, \rho_{ab}}{\sigma^2_{ab}}$} \\
\midrule
\textbf{Basis Decomposition} & \textbf{Pixel}~\cite{Pol:2022sjn} & \textbf{Spherical Harmonics}~\cite{Mingarelli:2013dsa,Pol:2022sjn} & \textbf{Principal Maps}~\cite{Ali-Haimoud:2020ozu,alihaimoud2021} \\
\midrule
True power $P(\hat{\Omega})$ 
 & $\sum_{p=1}^{N_{\rm pix}} P_p \, \delta^2(\hat{\Omega}-\hat{\Omega}_p)$ 
 & $\sum_{\ell=0}^{\lgwb} \sum_{m=-\ell}^{\ell} P_{\ell m} \, Y_{\ell m}(\hat{\Omega})$~\cite{Agarwal:2026nxa} 
 & $\sum_{n=1}^{N_{\rm pair}} P_n \, M_n(\hat{\Omega})$ \\
 \midrule
\multirow{1}{*}{\textbf{Regularized Method}} 
 &  Truncated SVD &  Truncated SVD & Choosing $N_{\rm modes}$ modes with highest eigenvalues \\
\midrule
\multirow{2}{*}{\textbf{Regularized Clean Map}} 
 & $\hat{P}_p = \sum_{n=1}^{N_{\rm modes}} M_{pn} \left[ \Sigma_n^{-1} M_{np'} X_{p'} \right]$ & $\hat{P}_{\ell m}=\sum_{n=1}^{N_{\rm modes}} M_{\ell m;n} \left[ \Sigma_n^{-1} M^*_{n;\ell'm'} X_{\ell'm'} \right]$ & $\hat{P}(\hatom)=\sum_{n=1}^{N_{\rm modes}} \hat{P}_n \, M_n(\hat{\Omega})$\\
& &$\hat{P}(\hatom)=\sum_{\ell m} \hat{P}_{\ell m} \, Y_{\ell m}(\hat{\Omega})$  &\\
 \midrule
Clean Map / ML Estimator of Coefficients 
 & $\hat{P}_p = (F^{-1})_{pp'} X_{p'}$ 
 & $\hat{P}_{\ell m} = (F^{-1})_{\ell m, \ell' m'} X_{\ell' m'}$ 
 & $\hat{P}_n = (F^{-1})_{nn'} X_{n'}$ \\
 PTA Basis Component & $\gamma_{ab;p}=\gamma_{ab}(\hatom_p)$& $\gamma_{ab;\ell m}=\int \d^2\hatom\,\gamma_{ab}(\hatom)\,Y_{\ell m}^*(\hatom)$ & $\gamma_{ab;n}=\int \d^2\hatom\,\gamma_{ab}(\hatom)\,M_n(\hatom)$\\
 Dirty Map & $X_p \equiv \sum_{ab} \frac{\gamma_{ab;p} \, \rho_{ab}}{\sigma^2_{ab}}$ & $X_{\ell m} \equiv \sum_{ab} \frac{\gamma^*_{ab;\ell m} \, \rho_{ab}}{\sigma^2_{ab}}$& $X_n \equiv \sum_{ab} \frac{\gamma_{ab;n} \, \rho_{ab}}{\sigma^2_{ab}}$\\
 Fisher Matrix & $F_{pp'} \equiv \sum_{ab} \frac{\gamma_{ab;p} \, \gamma_{ab;p'}}{\sigma^2_{ab}}$ & $F_{\ell m,\ell' m'}\equiv \sum_{ab} \frac{\gamma^*_{ab;\ell m}\,\gamma_{ab;\ell' m'}}{\sigma^2_{ab}}$ & $F_{nn'} \equiv \sum_{ab} \frac{\gamma_{ab;n} \, \gamma_{ab;n'}}{\sigma^2_{ab}}$ \\
 \midrule
Singular value decomposition 
 & $F_{pp'} = \sum_{n=1}^{N_{\rm pix}} M_{pn} \, \Sigma_n \, M_{np'}$ & $F_{\ell m, \ell'm'} = \sum_{n=1}^{N_{\rm SpH}} M_{\ell m;n} \, \Sigma_n \, M^*_{n;\ell'm'}$~\cite{Agarwal:2026nxa} & \\
Regularized Fisher Inverse 
 & $(F^+)_{pp'} = \sum_{n=1}^{N_{\rm modes}} M_{pn} \, \Sigma_n^{-1} \, M_{np'}$ & $(F^+)_{\ell m, \ell'm'} = \sum_{n=1}^{N_{\rm modes}} M_{\ell m;n} \, \Sigma^{-1}_n \, M^*_{n;\ell'm'}$ & \\
\midrule
\multirow{2}{*}{Noise Properties~\cite{Agarwal:2026nxa}} & \multicolumn{3}{c}{Mean$\quad \langle \bm{\hat{P}}\rangle =\bm{F}^+ \bm{F} \bm{P}$}\\
& \multicolumn{3}{c}{Covariance$[\bm{\hat{P}} ]=\bm{F}^+ \bm{F} \bm{F}^+$}\\
\bottomrule
\end{tabular}}
\caption{Summary of directional GWB power estimation in the weak signal limit, showing the relationship between crosscorrelations, the Fisher information matrix, and maximum likelihood map reconstruction in different bases. Importantly, the regularized clean map has an identical structure in the pixel, SpH and principal maps bases. In this article, we show that the numerical results are also identical when same the information content (i.e., the same set of reconstructed eigenmodes) are used in each basis.}
\label{tab:cleaned_gwb}
\end{table*}

We consider the frequentist formalism for anisotropy searches~\cite{Mitra:2007mc,Thrane:2009fp,Pol:2022sjn}, commonly implemented using pixel~\cite{Mitra:2007mc,Cornish:2014rva,Pol:2022sjn} and the spherical harmonic (SpH) bases~\cite{Thrane:2009fp,Mingarelli:2013dsa,Pol:2022sjn}. A Fisher matrix and principal maps (or eigenmaps) formulation based on the pairwise timing response (PTR) has also been developed by~\citet{Ali-Haimoud:2020ozu,alihaimoud2021}. All of these approaches require choosing a finite-dimensional truncation of the sky, together with a regularization scheme for inverting the Fisher matrix when necessary:
\begin{itemize}
    \item Pixel basis. The number of pixels used to decompose the sky and regularization parameter used to obtain pseudoinverse of Fisher information matrix (if required; computed via truncated singular value decomposition (SVD) here~\cite{Agarwal:2026nxa});
    \item SpH basis. The maximum multipole $\lrec$ used to truncate SpH expansion and regularization parameter used to obtain pseudoinverse of Fisher information matrix (if required; computed via truncated singular value decomposition (SVD) here);
    \item Principal maps basis. The pixelization/$\lrec$ used to construct the Fisher matrix in the PTR basis~\cite{Agarwal:2026nxa} (and the principal (eigen) maps) and number of eigenmodes used to reconstruct sky map (which is the same as truncated SVD regularization).
\end{itemize}
(See Table~\ref{tab:cleaned_gwb} for a summary of frequentist anisotropy reconstruction methods.)

In the PTA literature, these parameters in pixel and SpH searches are often chosen using a counting argument~\cite{Semenzato:2025sqc,Domcke:2025esw}. The idea is that, given $N_{\rm pair}$ independent pulsar-pair correlation measurements, one has access to $N_{\rm pair}$ independent pieces of information. This reasoning is then translated into the number of parameters used in the basis decomposition,
\begin{equation}
    12 \,N_{\rm side}^2=N_{\rm pair},\quad (\lNpair+1)^2=N_{\rm pair}
\end{equation}
which defines the pixel resolution on a {\tt HEALPix}\footnote{http://healpix.sf.net}~\cite{2005ApJ...622..759G,Zonca2019} grid and harmonic truncation, respectively. Another counting-based limit, originally derived for phase-coherent mapping~\cite{Romano:2016dpx}, has also been adopted for intensity mapping and applied to PTA data, $\lrec\sim \sqrt{N_{\rm psr}}$~\cite{Boyle:2010rt,Romano:2016dpx,Pol:2022sjn,NANOGrav:2023tcn}. 
This choice truncates the reconstruction at an even smaller multipole, thus potentially enhancing the artifacts.

However, this choice does not represent a fundamental information limit of the PTA. As shown by~\citet{Grunthal2026} in the context of point-source reconstruction using the SpH basis, the appropriate reconstruction multipole $\lrec$ need not coincide with $\lNpair$. In fact, using a higher $\lrec$ along with appropriate regularization scheme can yield a finer point-spread function. Moreover, the optimal truncation multipole depends on the pulsar sky configuration, not merely on $N_{\rm pair}$. 

Similarly, recent work~\cite{Semenzato:2025sqc} demonstrated that choosing $\lrec=\lNpair$ can lead to excess leakage from unmodeled but informative angular scales (referred to as ``small scales'' in Refs.~\cite{Semenzato:2025sqc,Mingarelli:2026}) in the reconstruction of the angular power spectrum. In our follow-up work~\cite{Agarwal:2026nxa}, we instead identified a maximum informative angular scale, $\lres$, defined by the requirement that the chosen truncation parameter $\lrec$ be sufficiently large to span the full observable space. As shown in Ref.~\cite{Agarwal:2026nxa}, for a fixed number of pulsars, smaller nearest-neighbor angular separations between pulsars empirically yield a higher $\lres$. More generally, $\lres$ depends on both pulsar geometry and noise levels through the Fisher matrix.

Adopting $\lrec=\lres$ removes leakage-induced bias, leaving only the bias associated with angular scales that are intrinsically suppressed by the PTA response. In other words, the recovered map and angular power spectrum are primarily influenced by information from scales $\ell\leq\lres$; modes beyond $\lres$ do not contribute significantly to the reconstruction~\cite{Agarwal:2026nxa}. This demonstrates that $\lrec$ should not be set based on $N_{\rm pair}$, but according to a criterion that reflects the angular scales to which the pulsar network is sensitive.

\begin{figure*}
    \centering
    \includegraphics[scale=0.7]{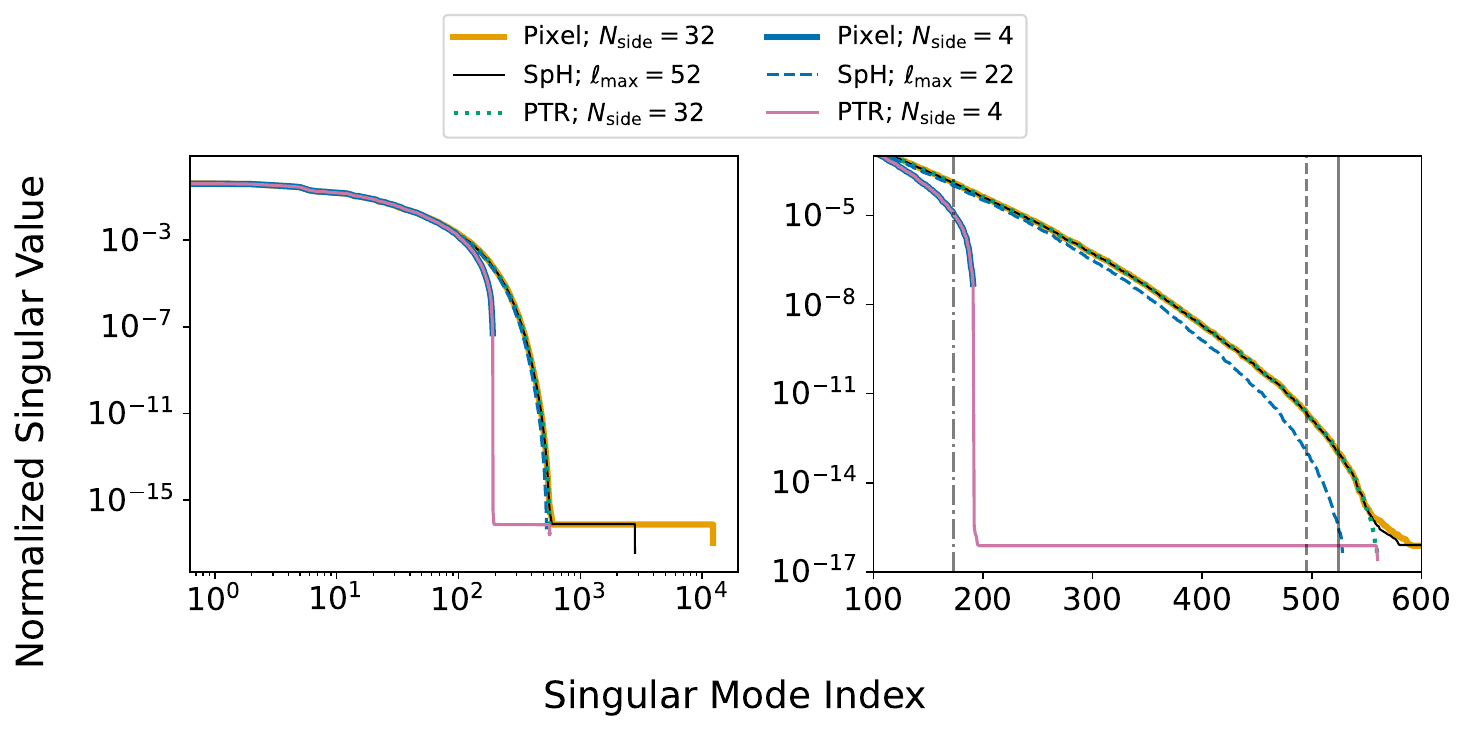}
    \caption{Singular value spectrum of the Fisher matrix in the pixel, spherical harmonic and pairwise timing response ($\bm{G}$; see Appendix~\ref{s:appendixA}) bases, shown for different choices of $N_{\rm side}$ and $\lrec$. The PTA configuration is the same as in Fig.~\ref{fig:pointSource_noNoise}. Left panel: truncation based on counting arguments (blue solid for pixel, blue dashed for SpH and pink solid for PTR) do not fully span the observable space, whereas higher-resolution choices---$N_{\rm side}$ = 32 for the pixel (orange solid) and PTR (green dotted) bases and $\lrec=52$ for the SpH basis (black solid)---yield a converged singular value spectrum that remains nearly unchanged with further increase in resolution~\cite{Agarwal:2026nxa}. Right panel: zoom-in of the small singular values. The vertical lines indicate the maximum singular mode index used in reconstruction for the eigenspace-converged setup (solid), pixel/PTR counting-based choice (dot-dashed) and SpH counting based choice (dashed).}
    \label{fig:SingularValue}
\end{figure*}

It has been suggested that the Fisher matrix formalism developed in Refs.~\cite{Ali-Haimoud:2020ozu,alihaimoud2021} can mitigate issues such as small-scale leakage and provide advantage over SpH representations~\cite{Mingarelli:2026}. Here, we clarify the origin of these differences. The Fisher matrix is the fundamental object that determines anisotropy information in any basis, including pixel~\cite{Pol:2022sjn,NANOGrav:2023tcn} and SpH bases~\cite{Pol:2022sjn,Grunthal:2024sor,Grunthal2026}. All anisotropy reconstructions can be understood as projections onto the eigenmodes of this Fisher matrix. Principal maps formalism in Refs.~\cite{Ali-Haimoud:2020ozu,alihaimoud2021} make this structure explicit, but these maps do not contain information beyond that encoded in the Fisher matrix. Any apparent difference between bases arises only if the Fisher eigenspace is not fully resolved. In that case, different choices of truncation (the pixel number, or $\lmax$) effectively retain different subsets of information, which can appear as a basis-dependent effect (see Fig.~\ref{fig:SingularValue}). 

Crucially, in the implementation of Refs.~\cite{Ali-Haimoud:2020ozu,alihaimoud2021}, a sufficiently high-resolution pixelization is adopted to construct their $F_{IJ}$ matrix [see Ref. \cite{Ali-Haimoud:2020ozu}, Eq. (84)]---chosen based on convergence of the eigenvalues/eigenmaps rather than on $N_{\rm pair}$ counting---and only $N_{\rm pair}$ principal maps are constructed. This procedure ensures that the finest angular information is preserved, while if $N_{\rm pix}$ or $\lrec$ is solely determined by counting arguments, the eigenspace may not be fully converged~\cite{Agarwal:2026nxa}, and reconstructions could effectively use different amounts of information, which could misleadingly appear as a basis-dependent effect. An alternative way to view this is that reconstruction using the formalism of Refs.~\cite{Ali-Haimoud:2020ozu,alihaimoud2021} is not immune to small-scale leakage if pixelization used to construct the principal maps is too coarse to resolve the angular scales to which the PTA is sensitive. In this case, the reconstruction is subject to the same truncation effect that can affect any basis representation, including the SpH basis (see our simulations).

\begin{figure}[t]
    \centering
    \includegraphics[scale=0.2]{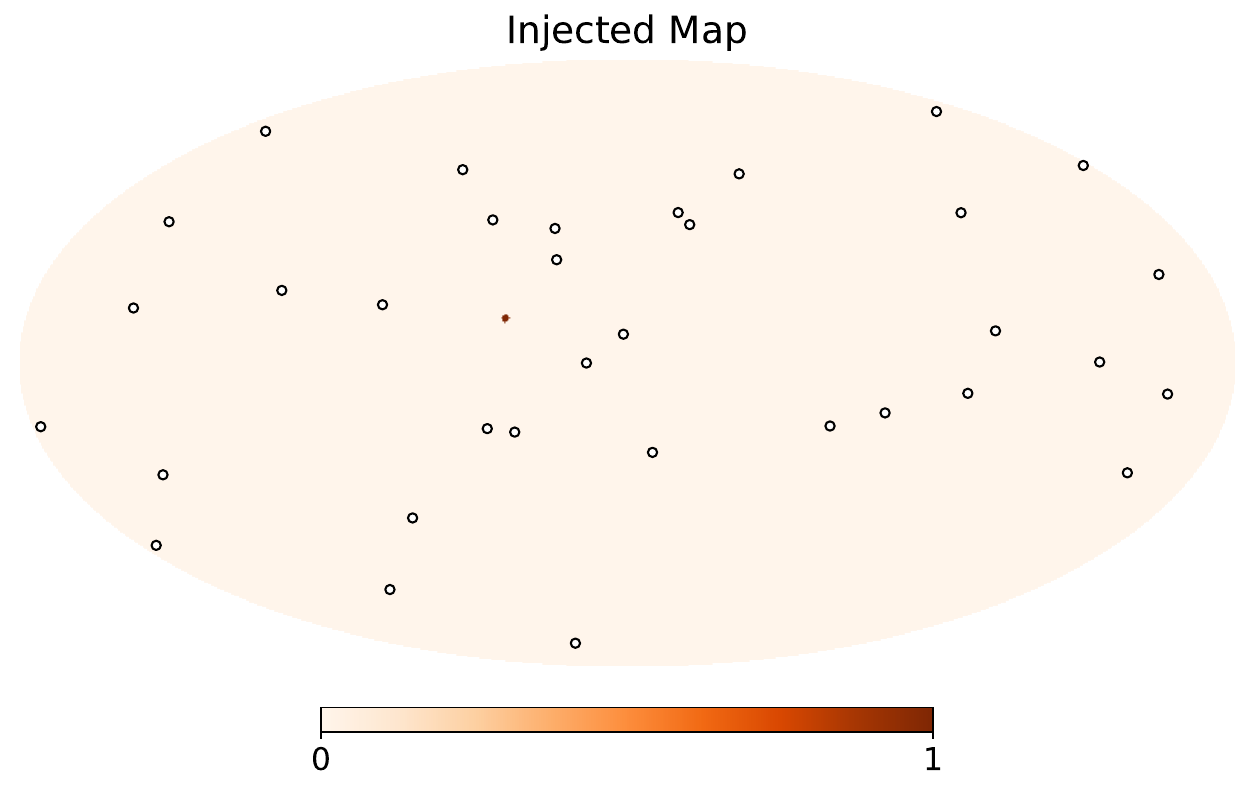} \includegraphics[scale=0.2]{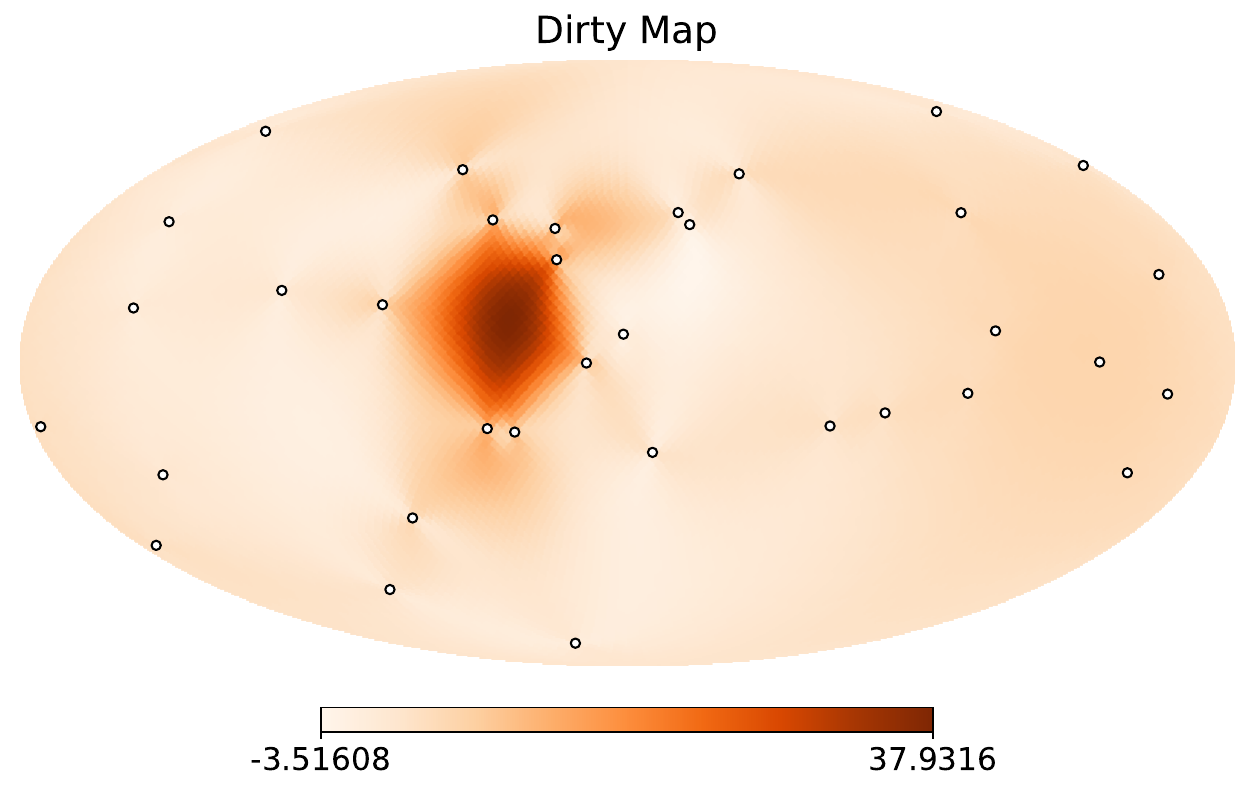}\\
    \includegraphics[scale=0.2]{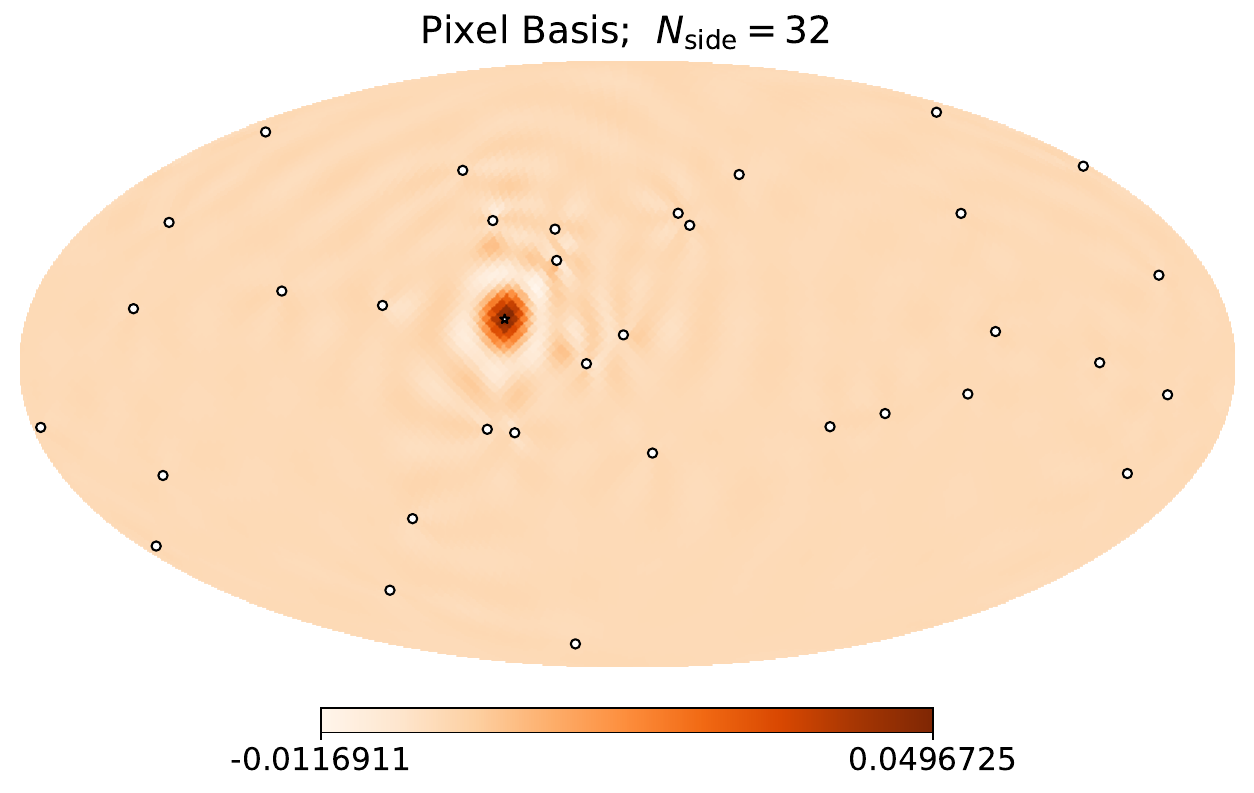} \includegraphics[scale=0.2]{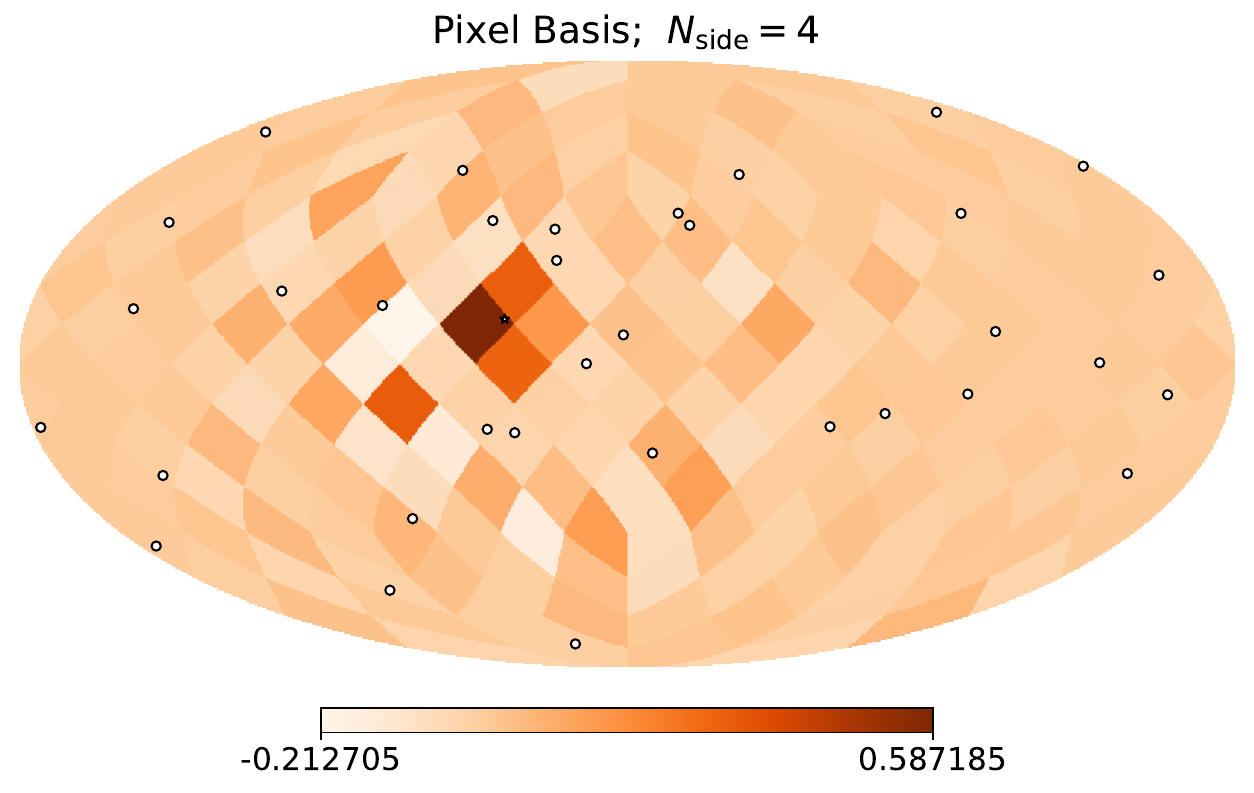}\\
    \includegraphics[scale=0.2]{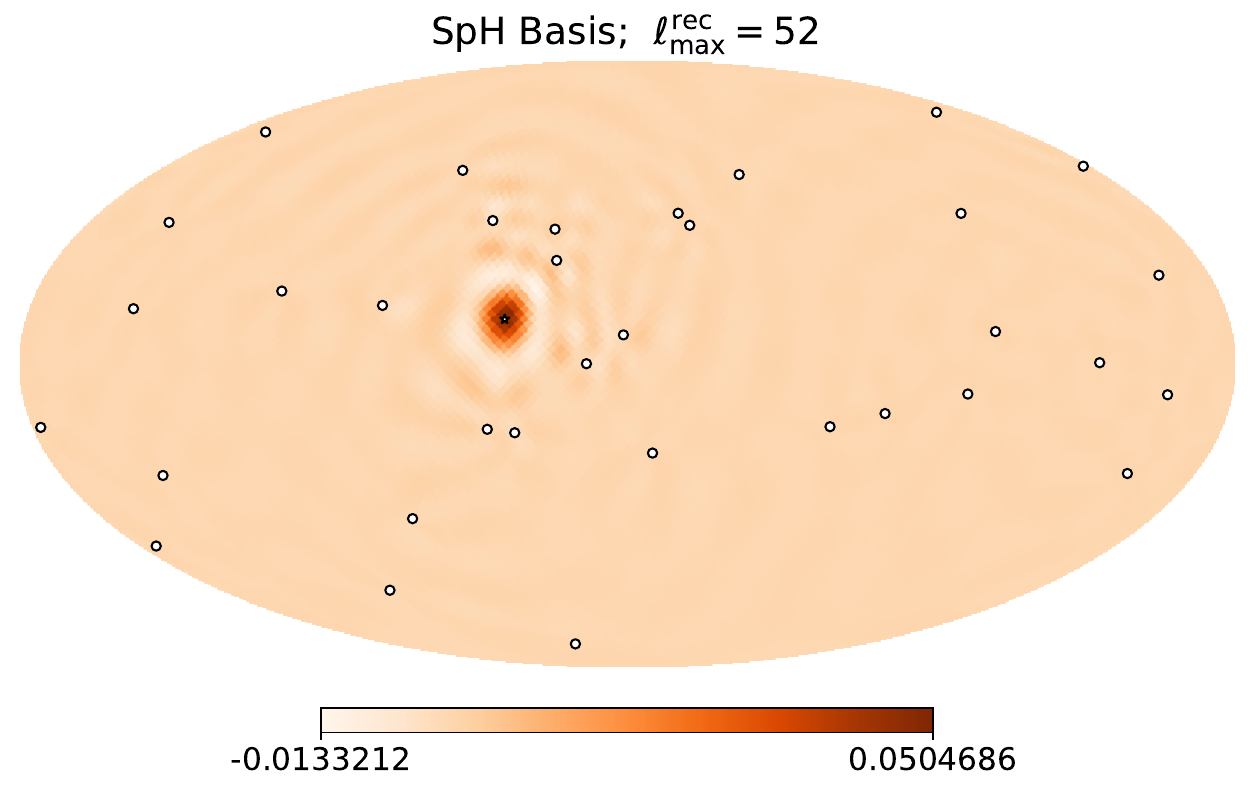} \includegraphics[scale=0.2]{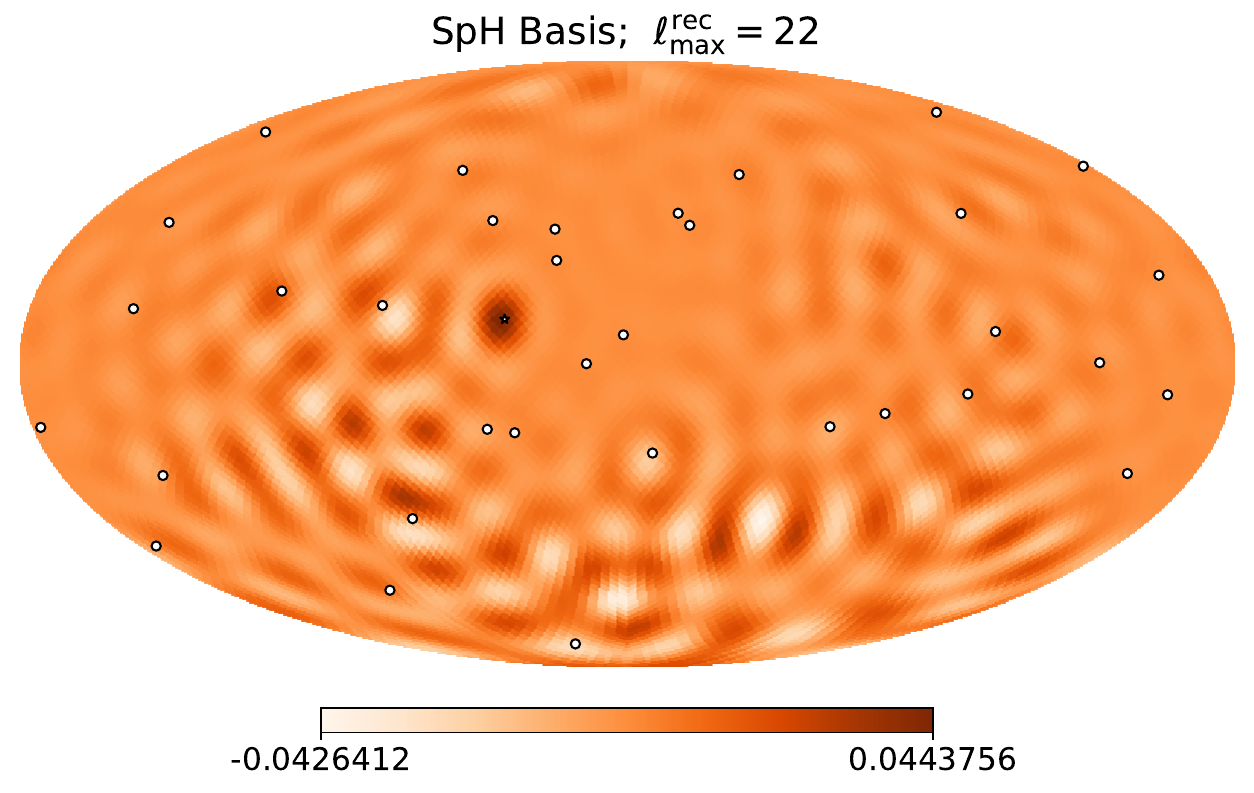}\\
    \includegraphics[scale=0.2]{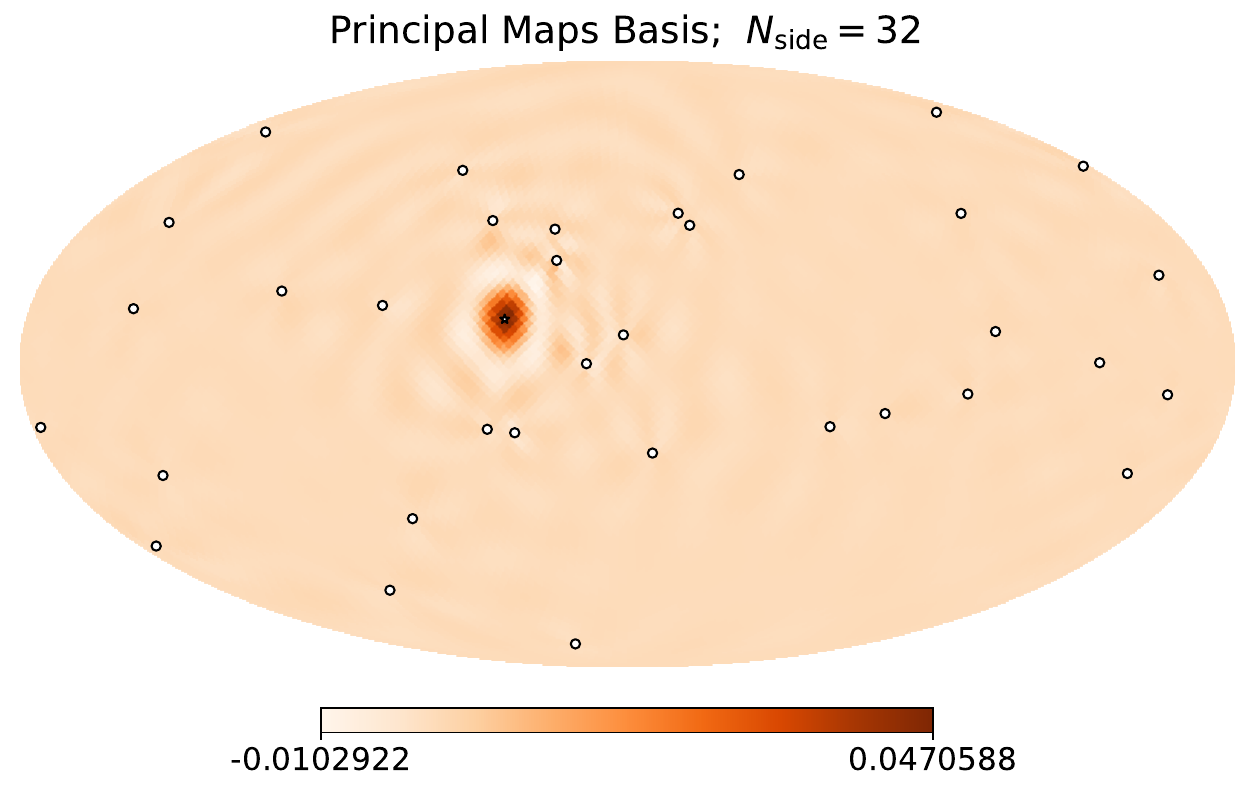}
    \includegraphics[scale=0.2]{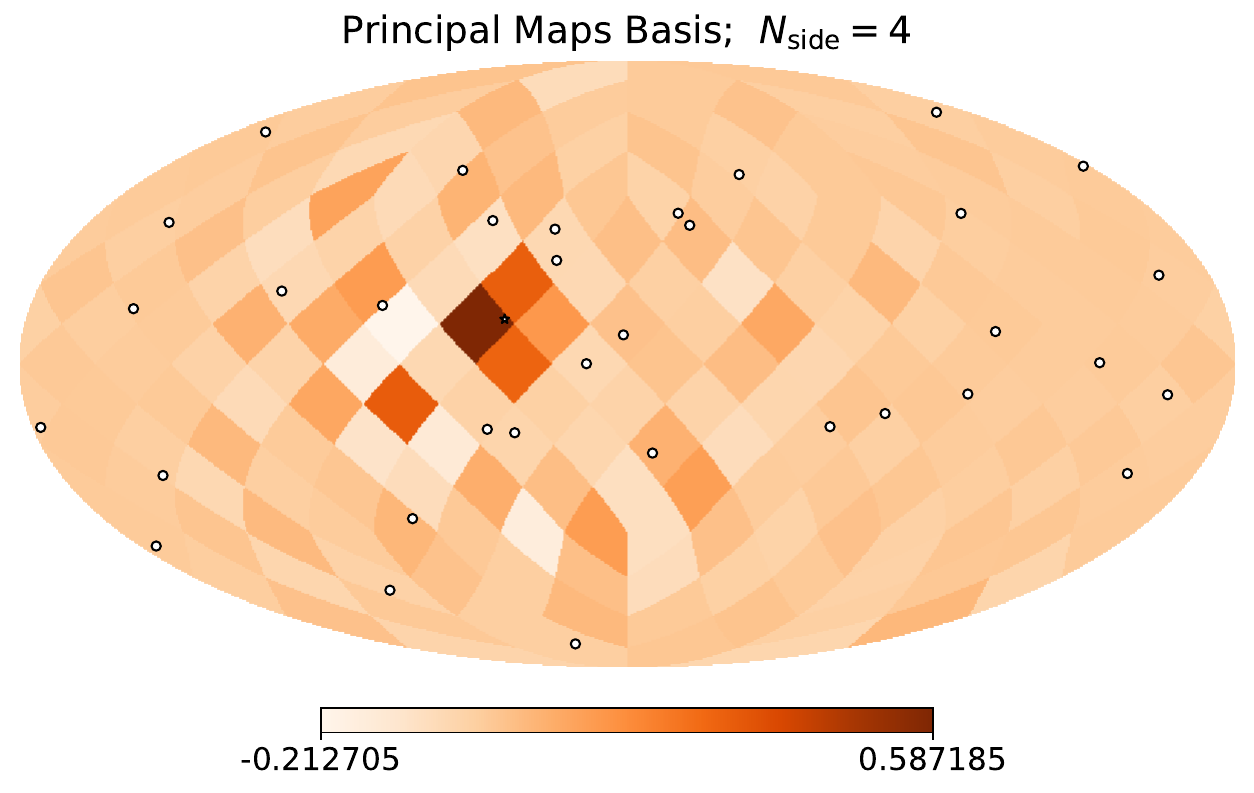}
    \caption{Reconstructed sky maps of the power $P_p$ (in units of strain$^2$) for a point source in the absence of noise ($\sigma_{ab}=0$). The top row shows the injected power and the corresponding dirty map in the pixel basis with $N_{\rm side}=32$. The left panels in the next three rows display reconstructions in the pixel, SpH and principal maps bases using eigenspace-converged setup and using a condition number threshold $\kappa=10^{13}$ (vertical solid line in Fig.~\ref{fig:SingularValue}). The agreement across these bases demonstrates their equivalence. Although power spreads across neighboring pixels, integrating regions with $P>0.5\, P_{\rm max}$ recovers 0.86 of the injected power. The right panel show reconstructions obtained using parameters determined from the counting argument and using $\kappa=10^5$ (dot-dashed vertical line in Fig.~\ref{fig:SingularValue}) for pixel and principal maps bases and $\kappa=10^{13}$ for SpH basis (dashed vertical line in Fig.~\ref{fig:SingularValue}).}
    \label{fig:pointSource_noNoise}
\end{figure}

\begin{figure}
    \centering
    \includegraphics[scale=0.2]{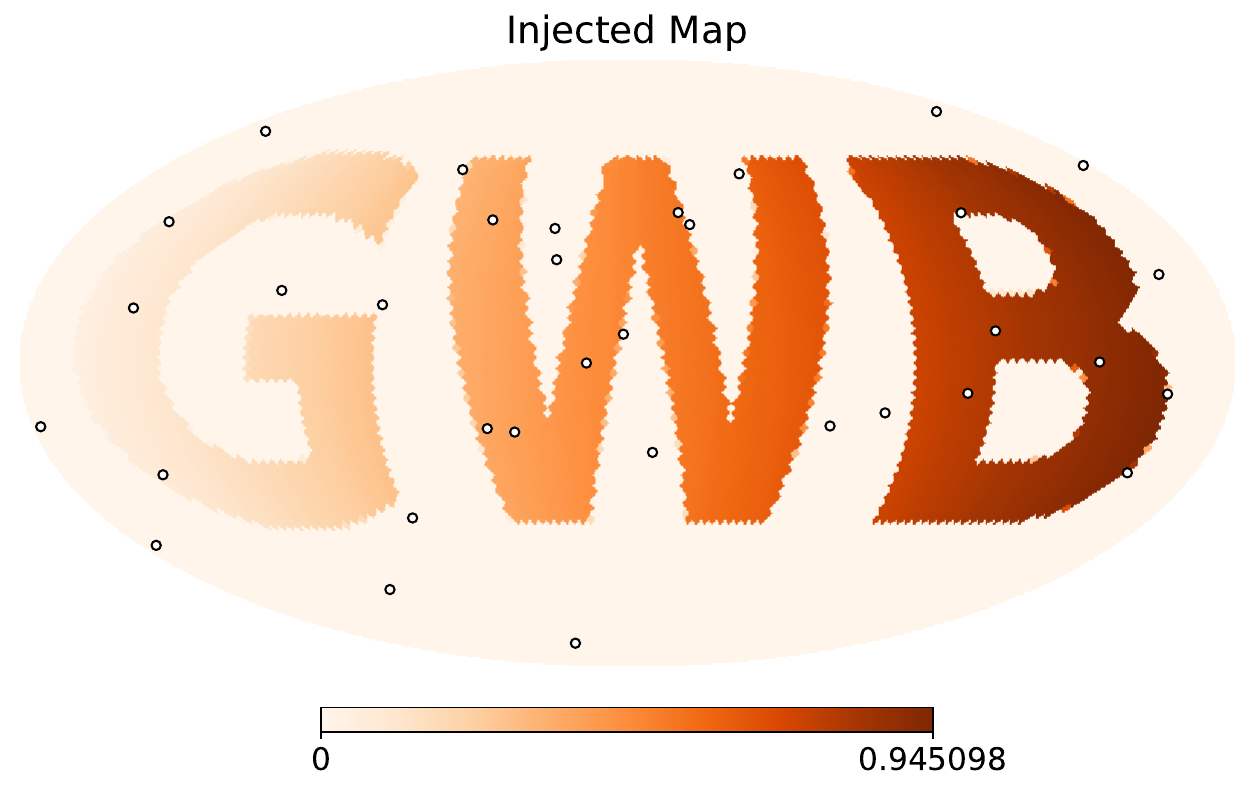}
    \includegraphics[scale=0.2]{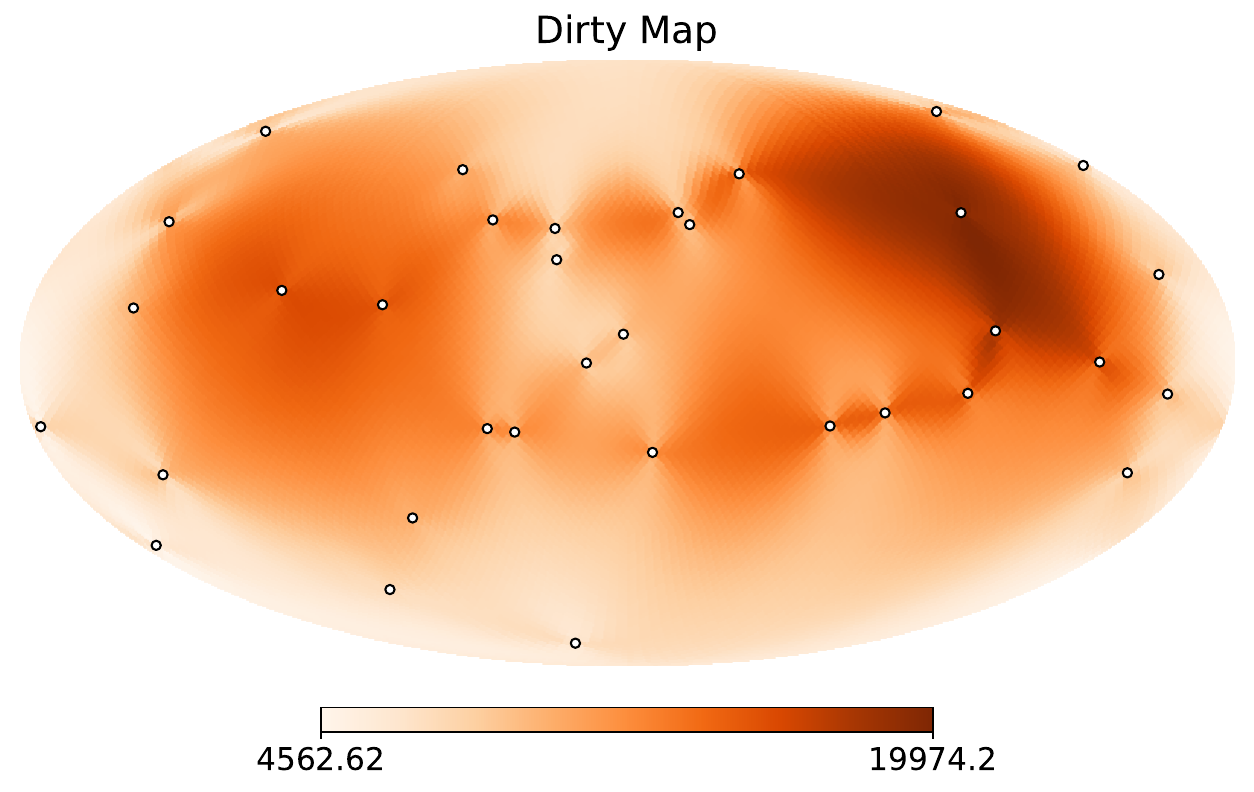}\\
    \includegraphics[scale=0.2]{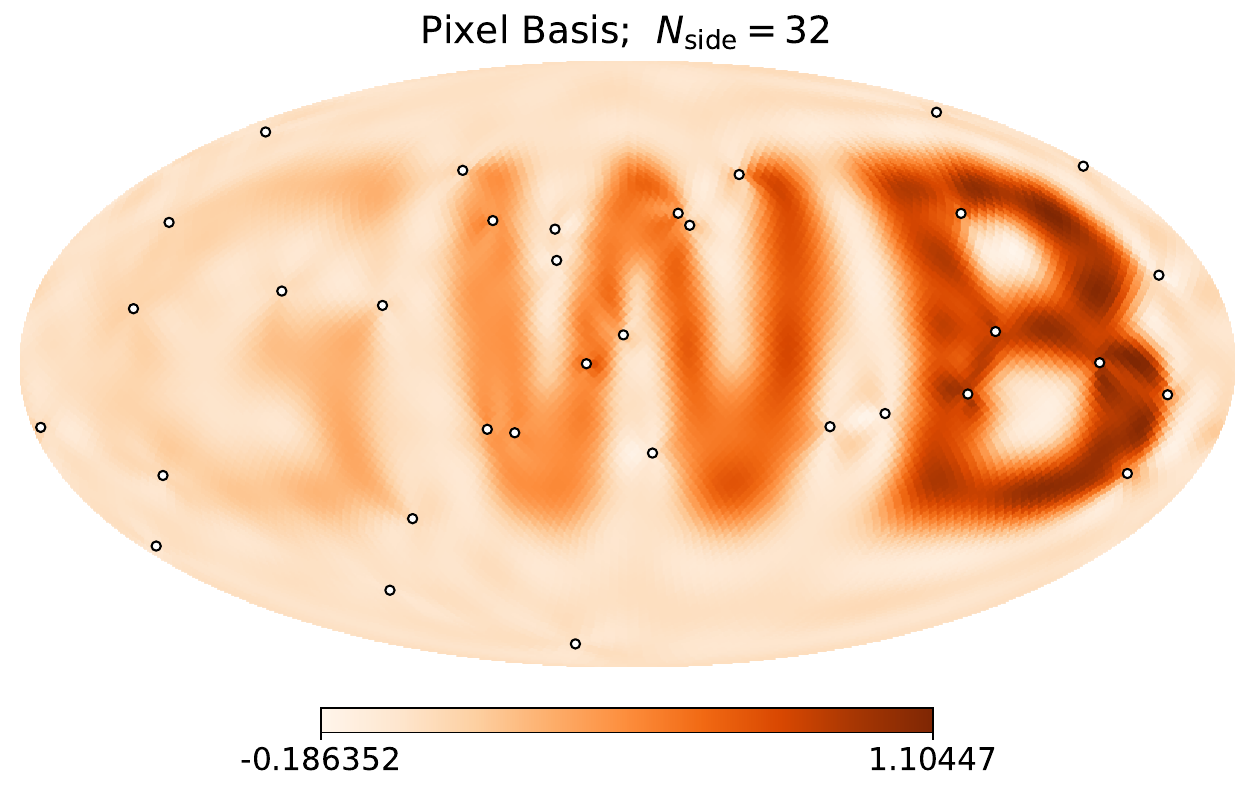}  \includegraphics[scale=0.2]{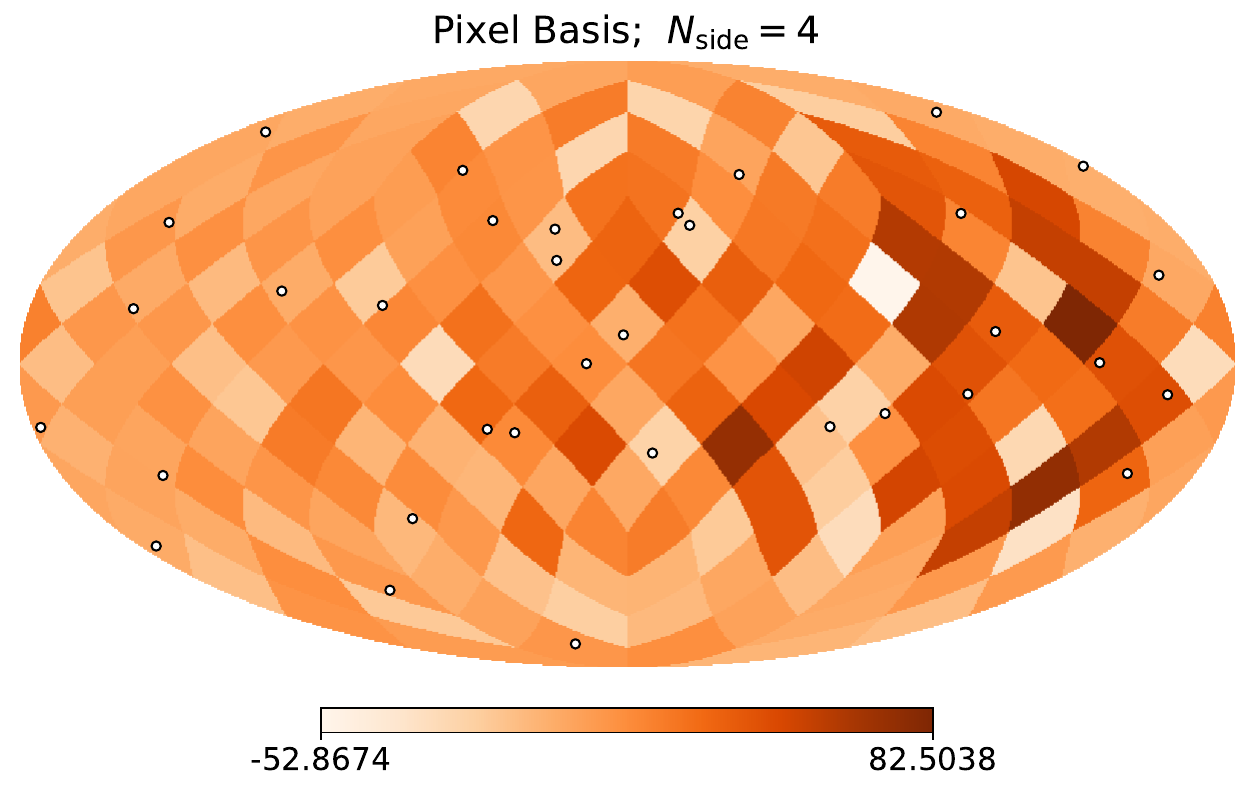}\\
    \includegraphics[scale=0.2]{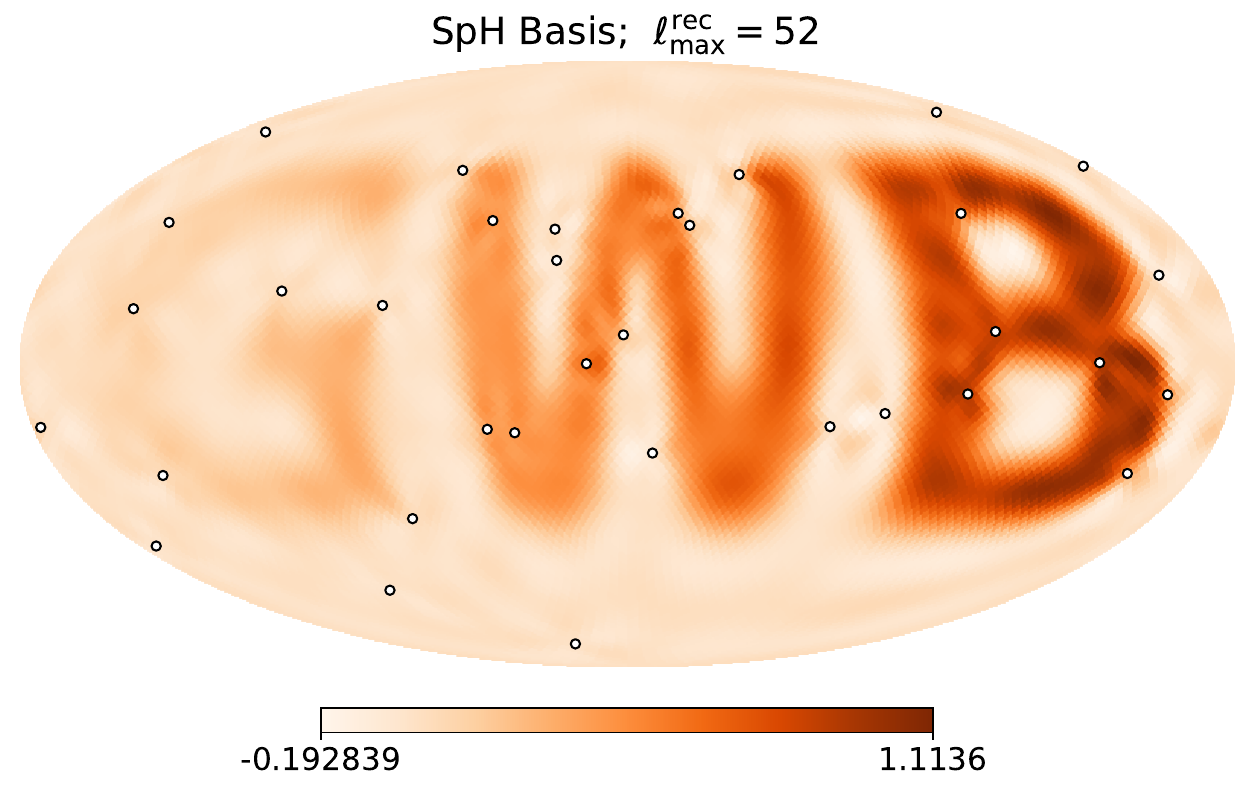}\includegraphics[scale=0.2]{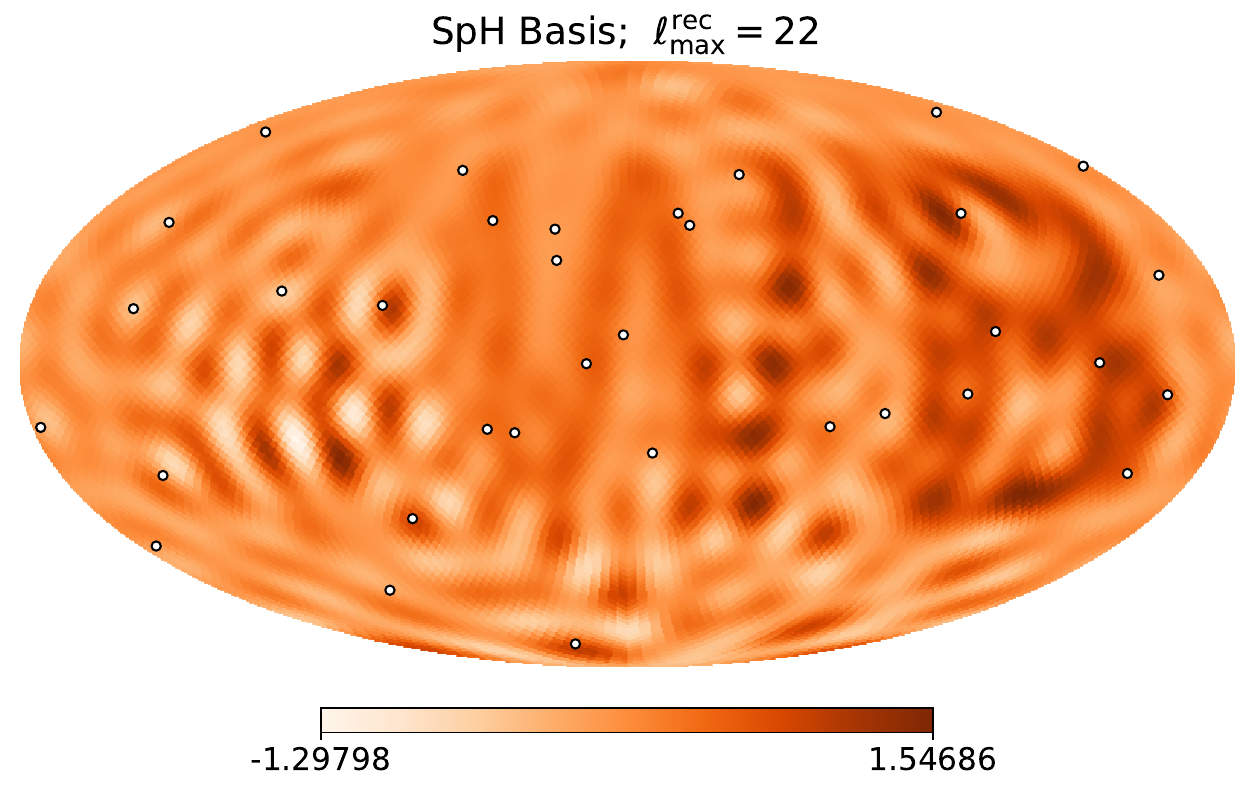}\\
    \includegraphics[scale=0.2]{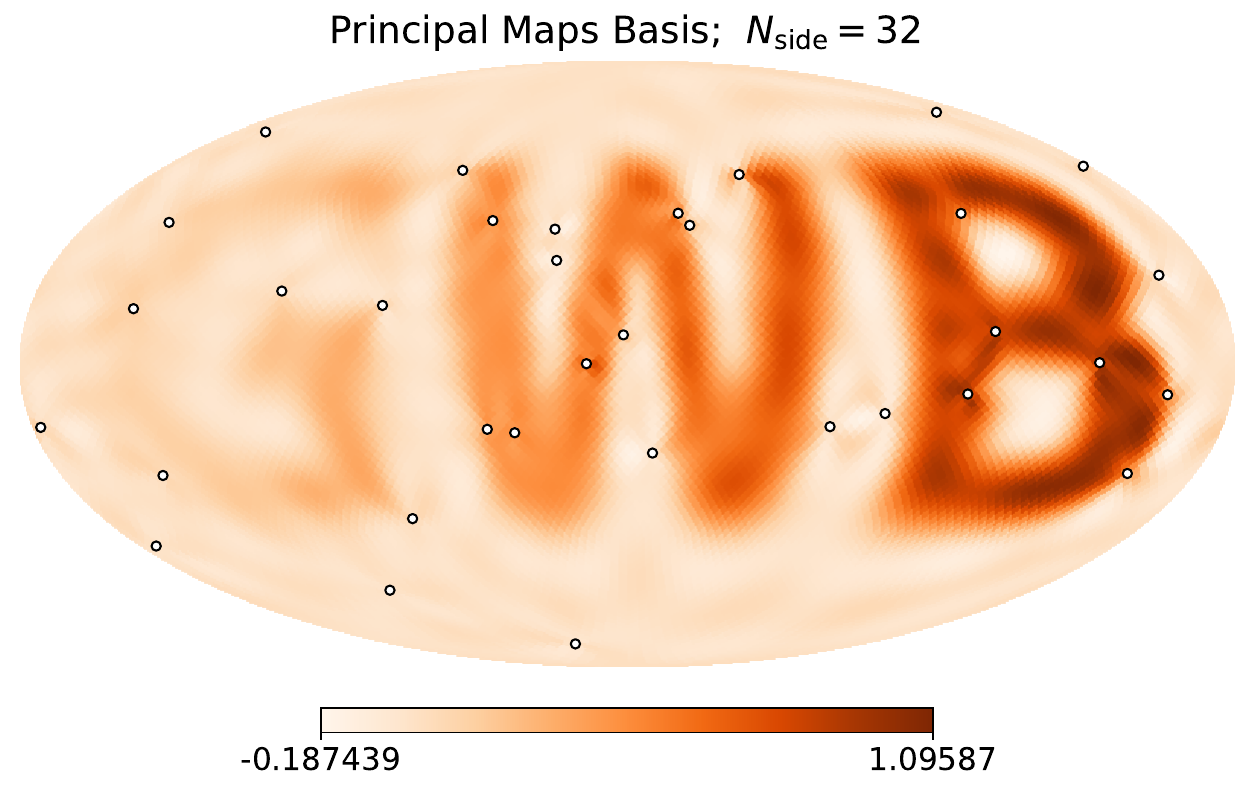}
    \includegraphics[scale=0.2]{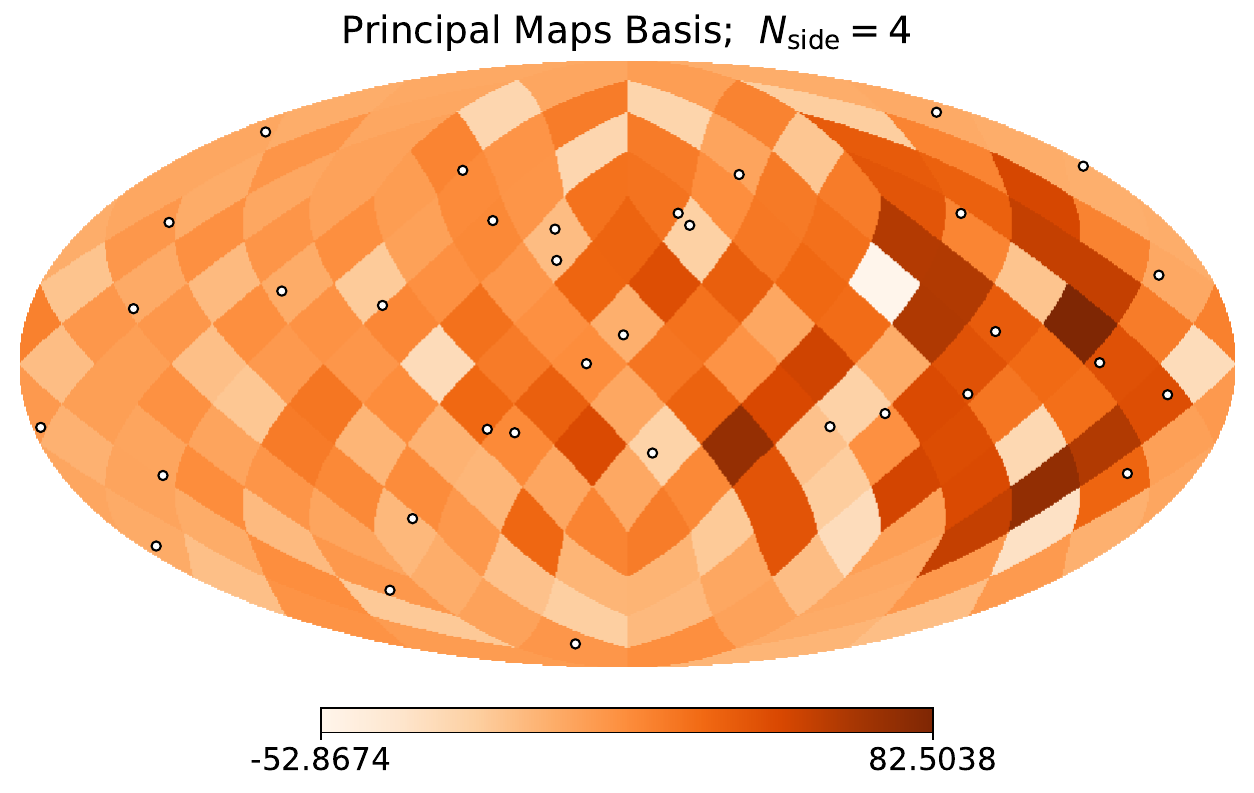}
    \caption{Reconstructed sky maps of the power $P_p$ (in units of strain$^2$) for a source with deterministic anisotropy in the absence of noise ($\sigma_{ab}=0$). The layout follows the same arrangement as Fig.~\ref{fig:pointSource_noNoise}.}
    \label{fig:extendedSource_noNoise}
\end{figure}

\begin{figure}
    \centering
    \includegraphics[scale=0.2]{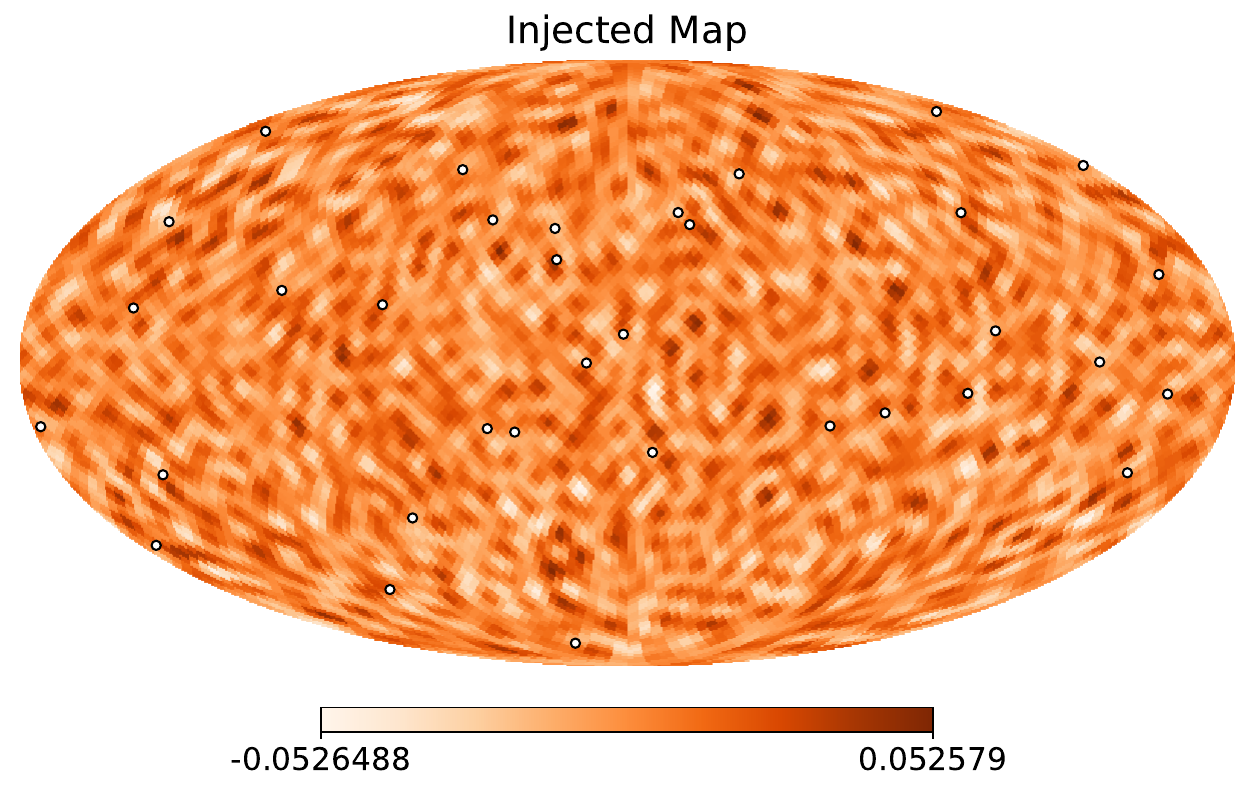}
    \includegraphics[scale=0.2]{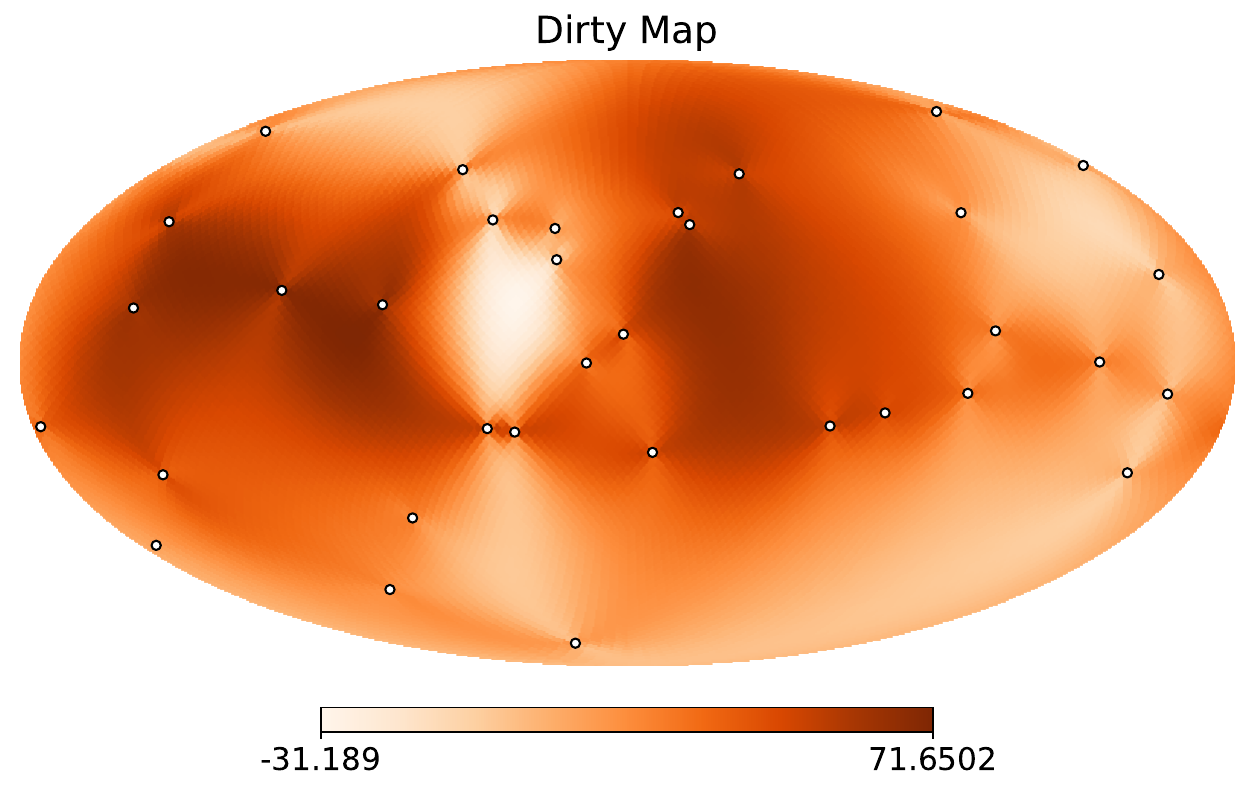}\\
    \includegraphics[scale=0.2]{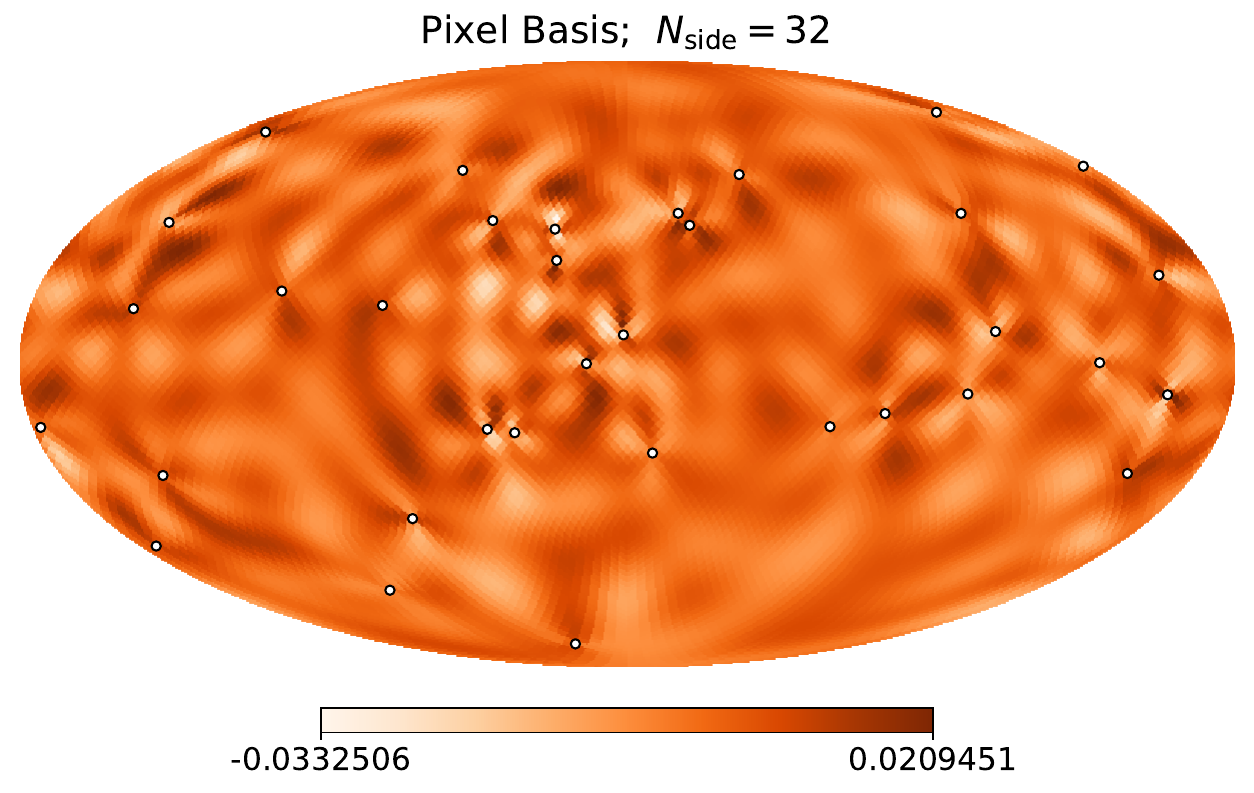}  \includegraphics[scale=0.2]{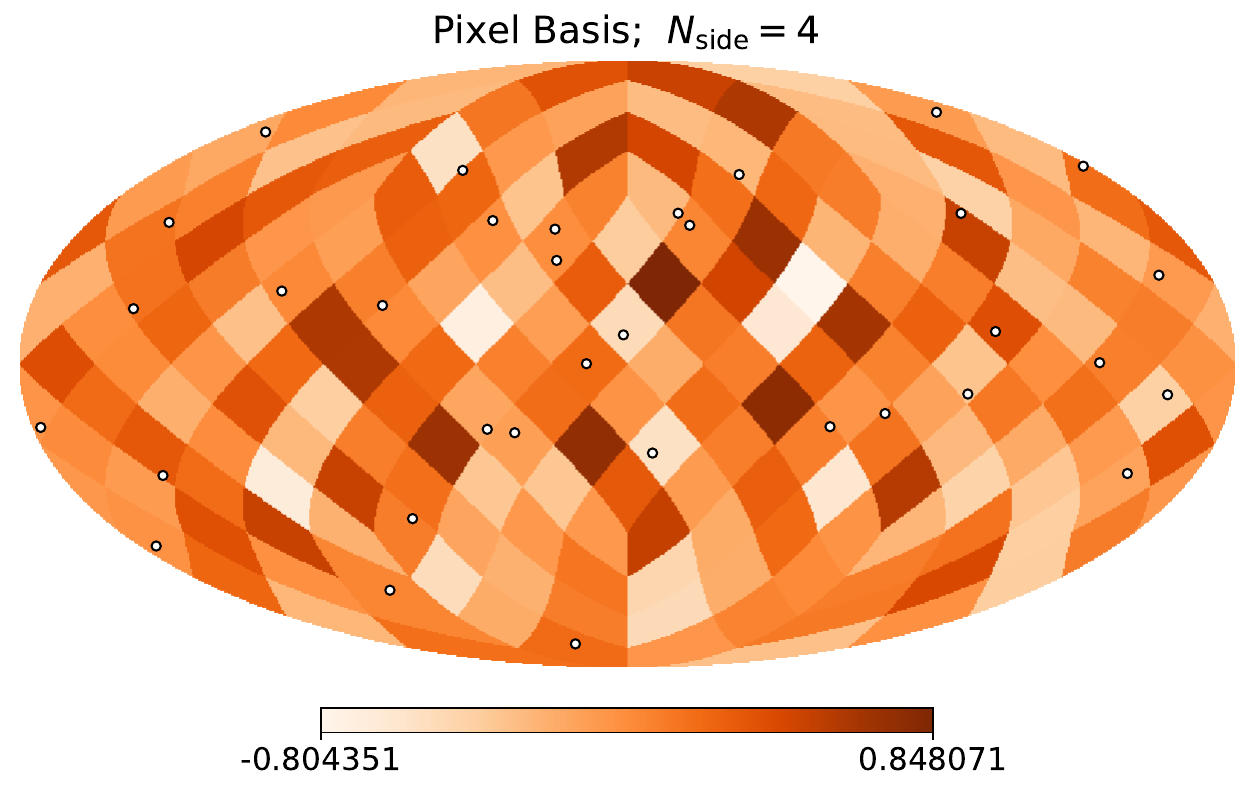}\\
    \includegraphics[scale=0.2]{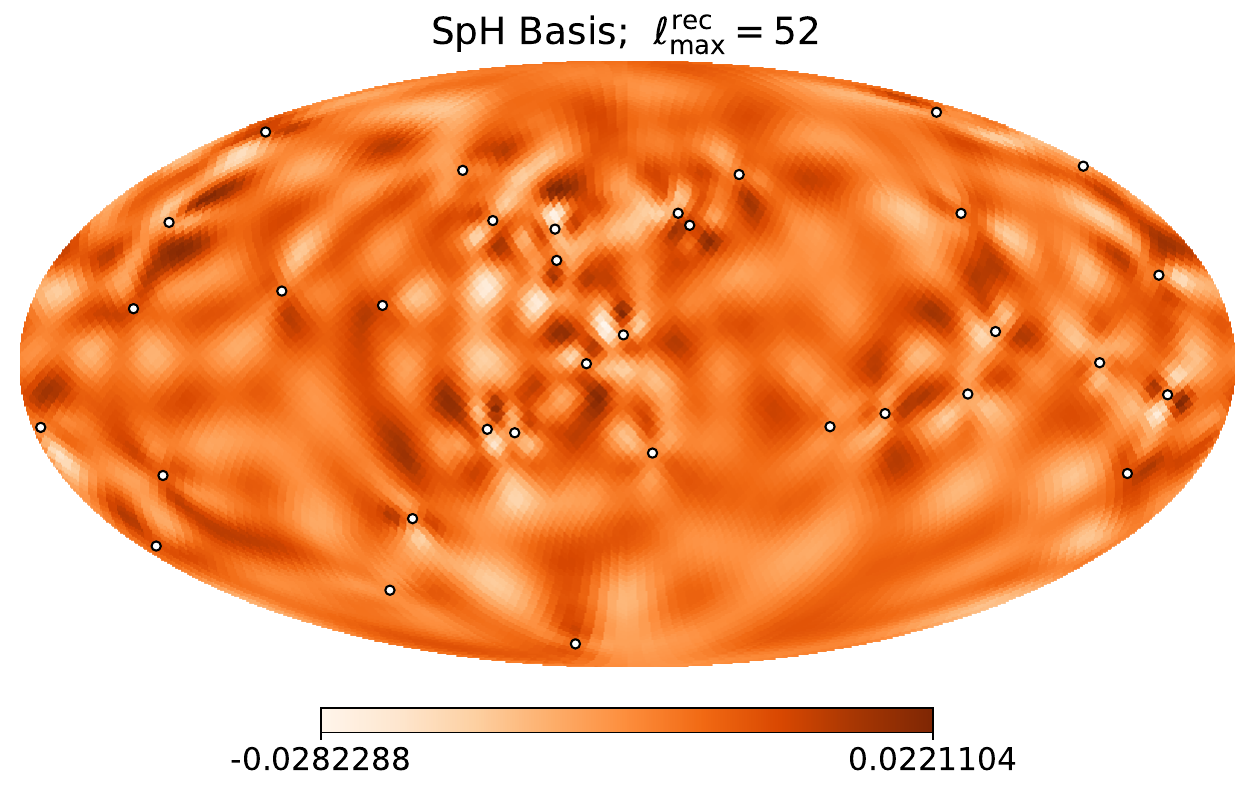}\includegraphics[scale=0.2]{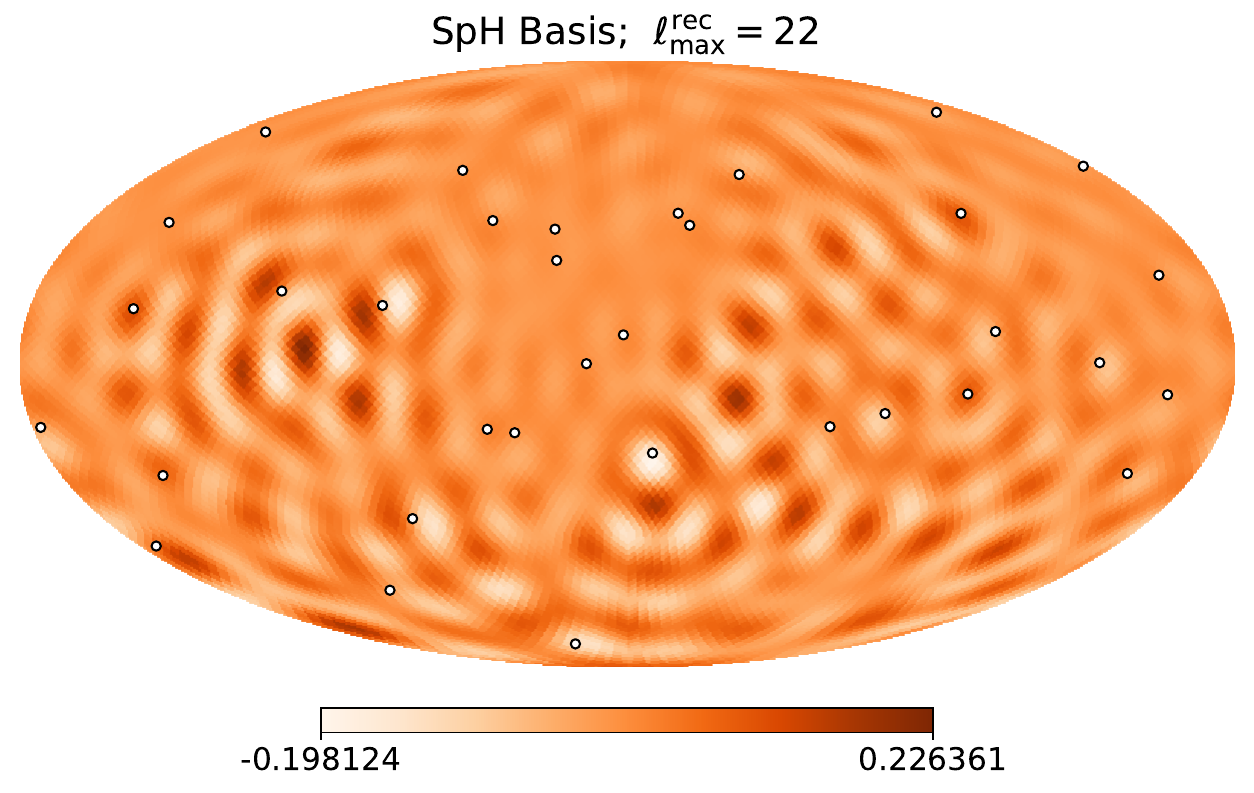}\\
    \includegraphics[scale=0.2]{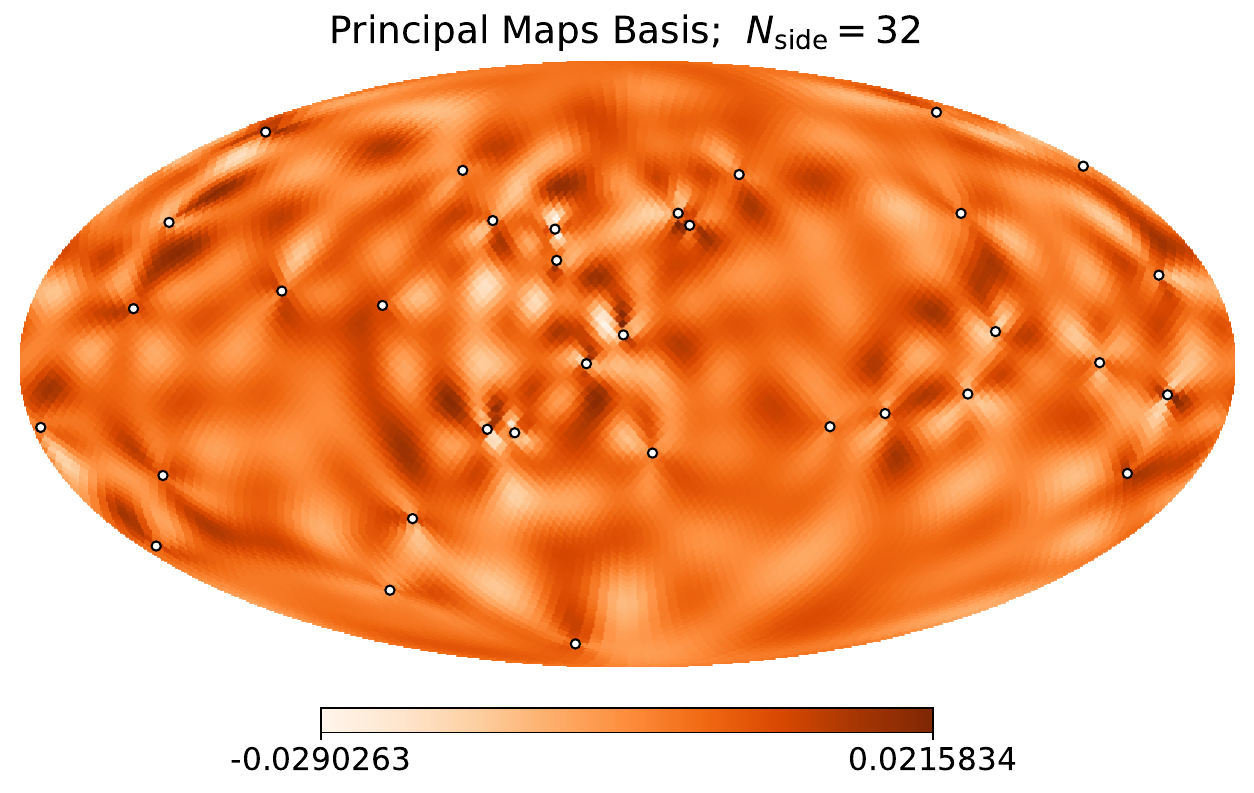}
    \includegraphics[scale=0.2]{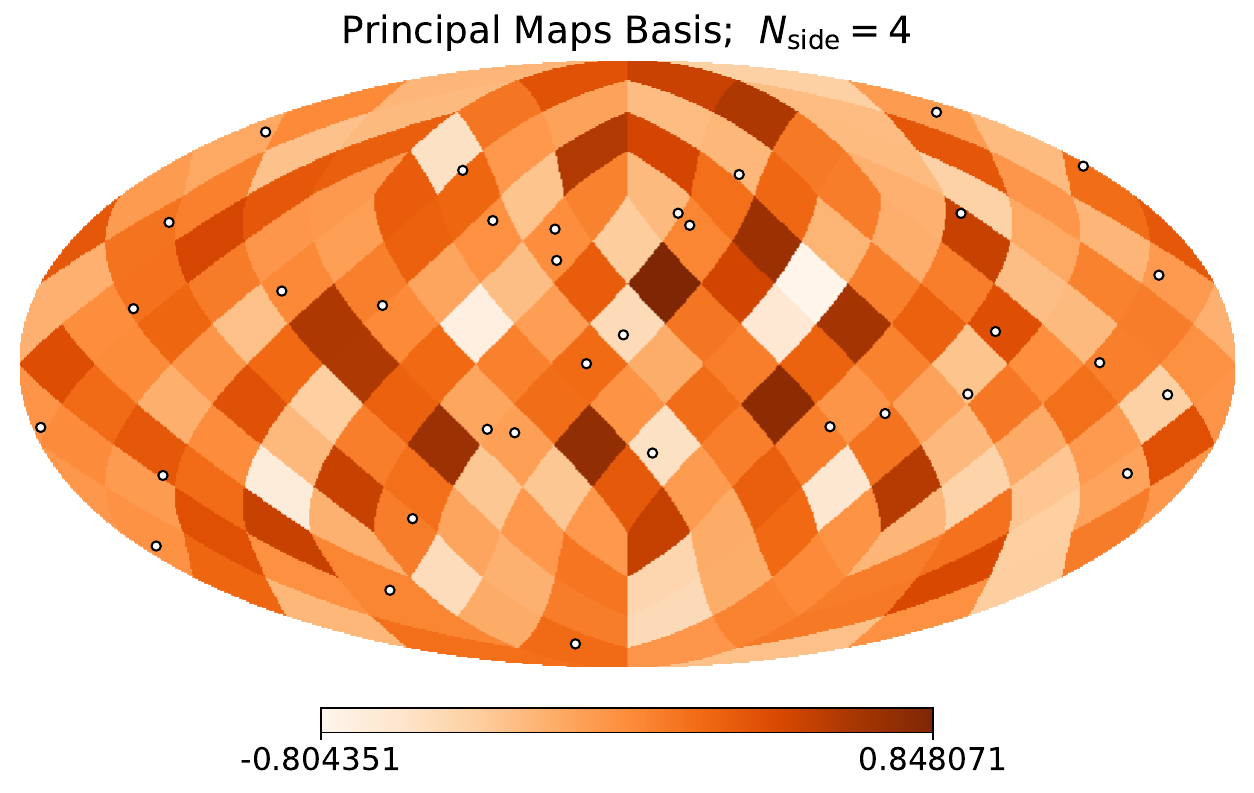}\\
    \includegraphics[scale=0.2]{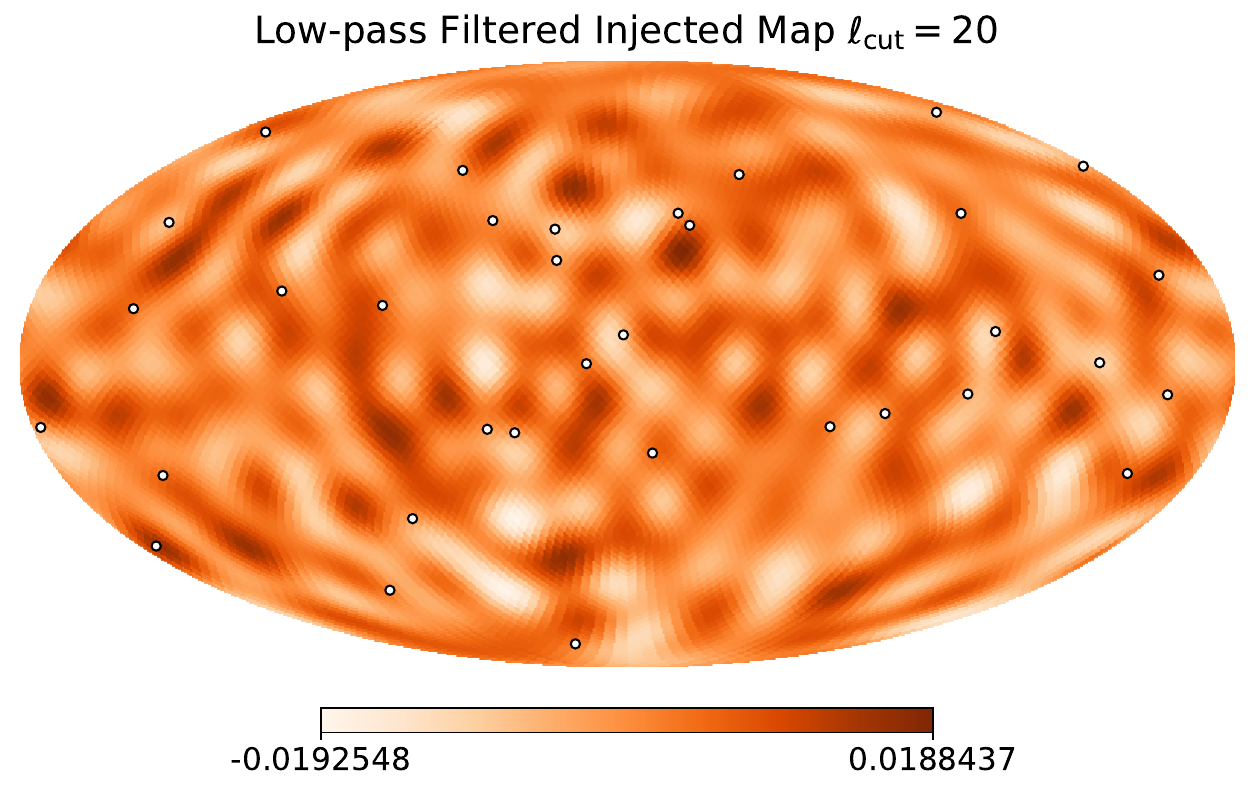}
    \caption{Reconstructed sky maps of the power $P_p$ (in units of strain$^2$) for a statistically isotropic source in the absence of noise ($\sigma_{ab}=0$). Layout follows the same arrangement for top four rows as in Fig.~\ref{fig:pointSource_noNoise}. The last row displays a low-pass filtered version of the injected map ($\ell_{\rm cut}=20$). Recovered maps in the left panels closely match these large-scale features, while finer details appear mainly in regions with denser pulsar coverage. The corresponding angular power spectrum recovery is shown in Fig.~\ref{fig:clRec_noNoise}.}
    \label{fig:statsIso_noNoise}
\end{figure}

\begin{figure}
    \centering
    \includegraphics[width=0.215\textwidth]{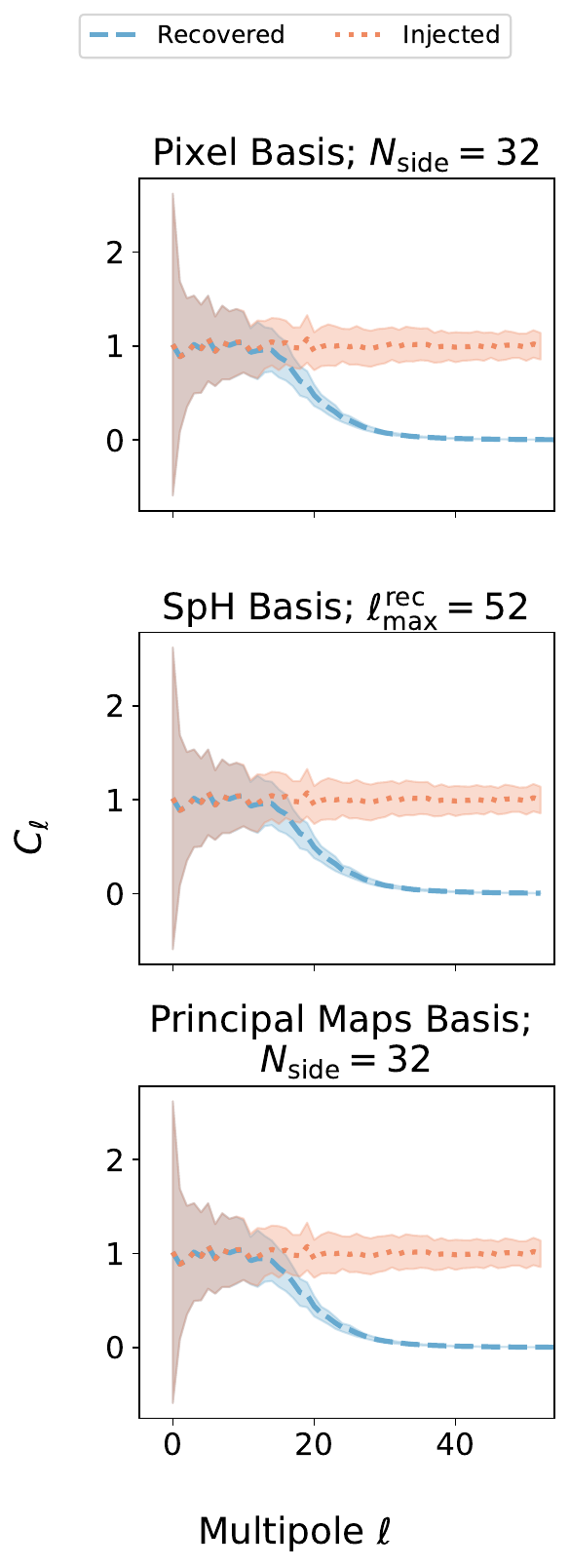}~~
    \includegraphics[width=0.22\textwidth]{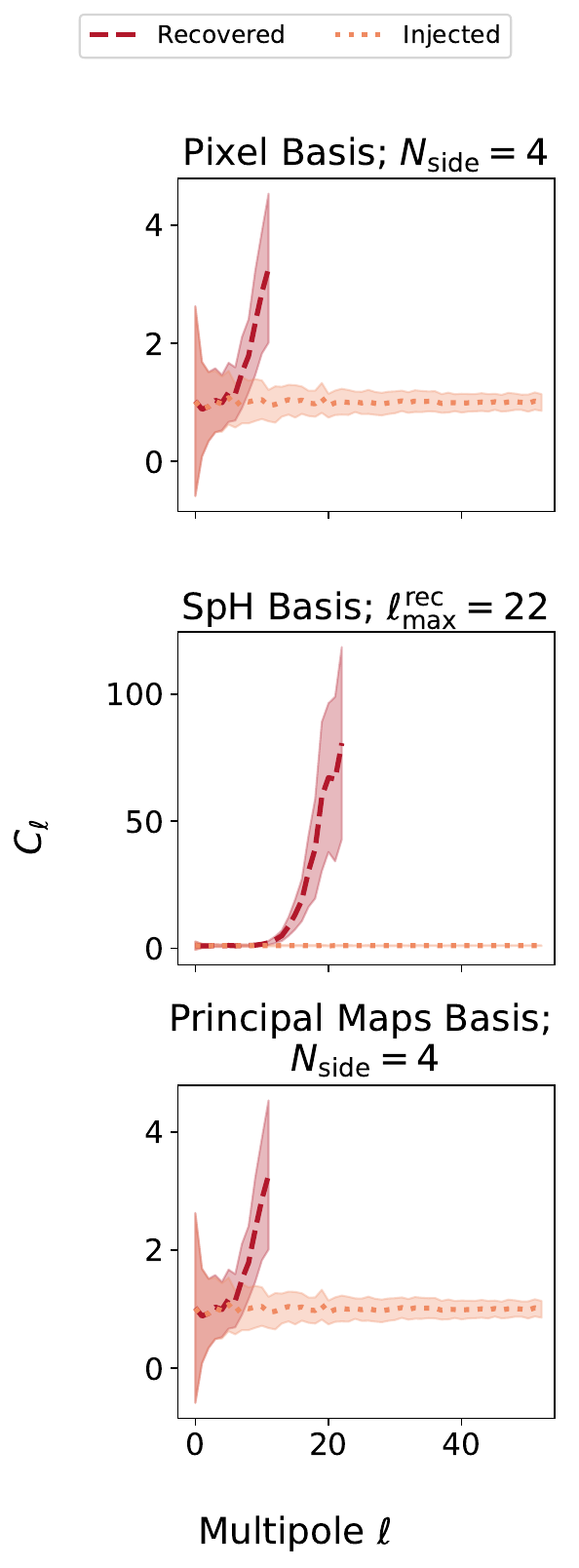}
    \caption{Angular power spectrum reconstruction without noise. The injected GWB power has support up to scale $\lgwb=52$, with the cosmic mean and 1$\sigma$ cosmic uncertainty from (100) simulated skies shown by orange dotted curve~\cite{Agarwal:2026nxa}. The layout follows the same arrangement as Fig.~\ref{fig:pointSource_noNoise} (from the second row). Recovery is consistent across bases in the left panels, whereas counting-argument-based parameters leads to leakage-induced bias~\cite{Semenzato:2025sqc,Agarwal:2026nxa}, independent of basis. }
    \label{fig:clRec_noNoise}
\end{figure}

\begin{figure}[htbp]
    \centering
   \includegraphics[scale=0.2]{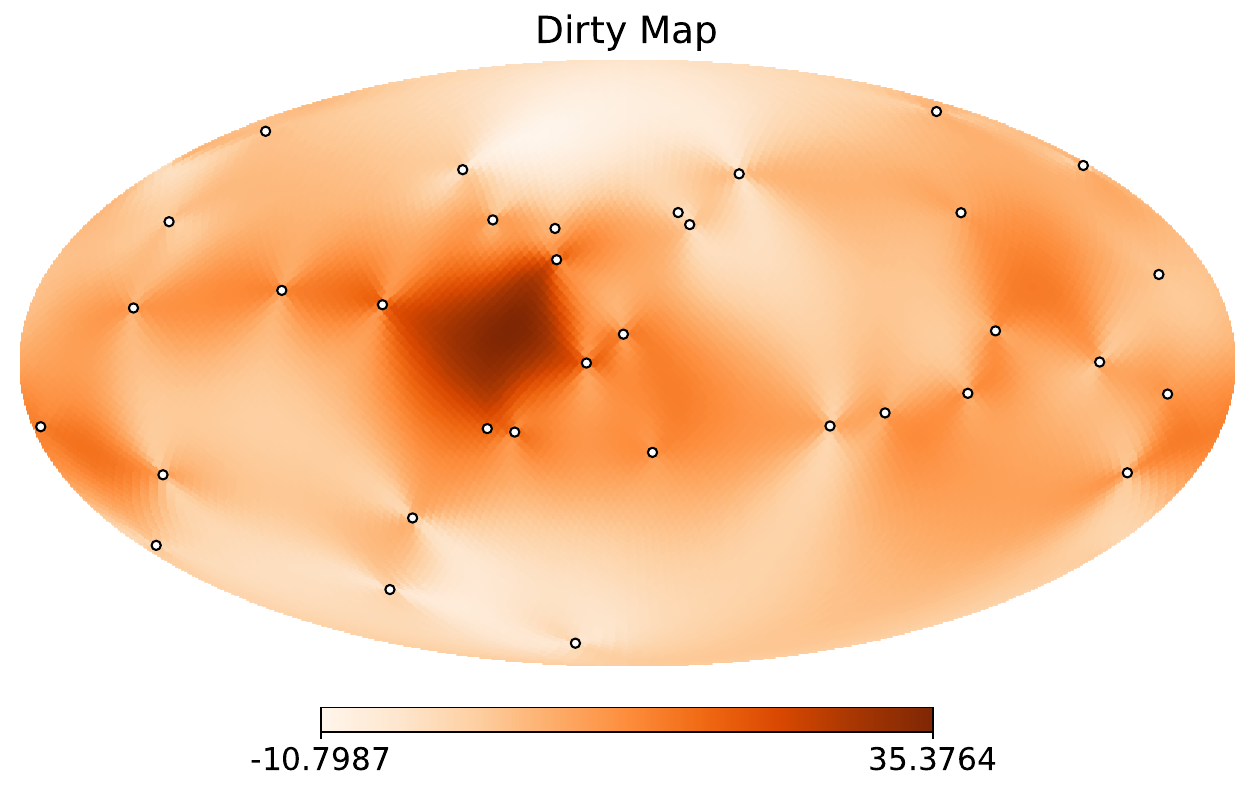}
    \includegraphics[scale=0.2]{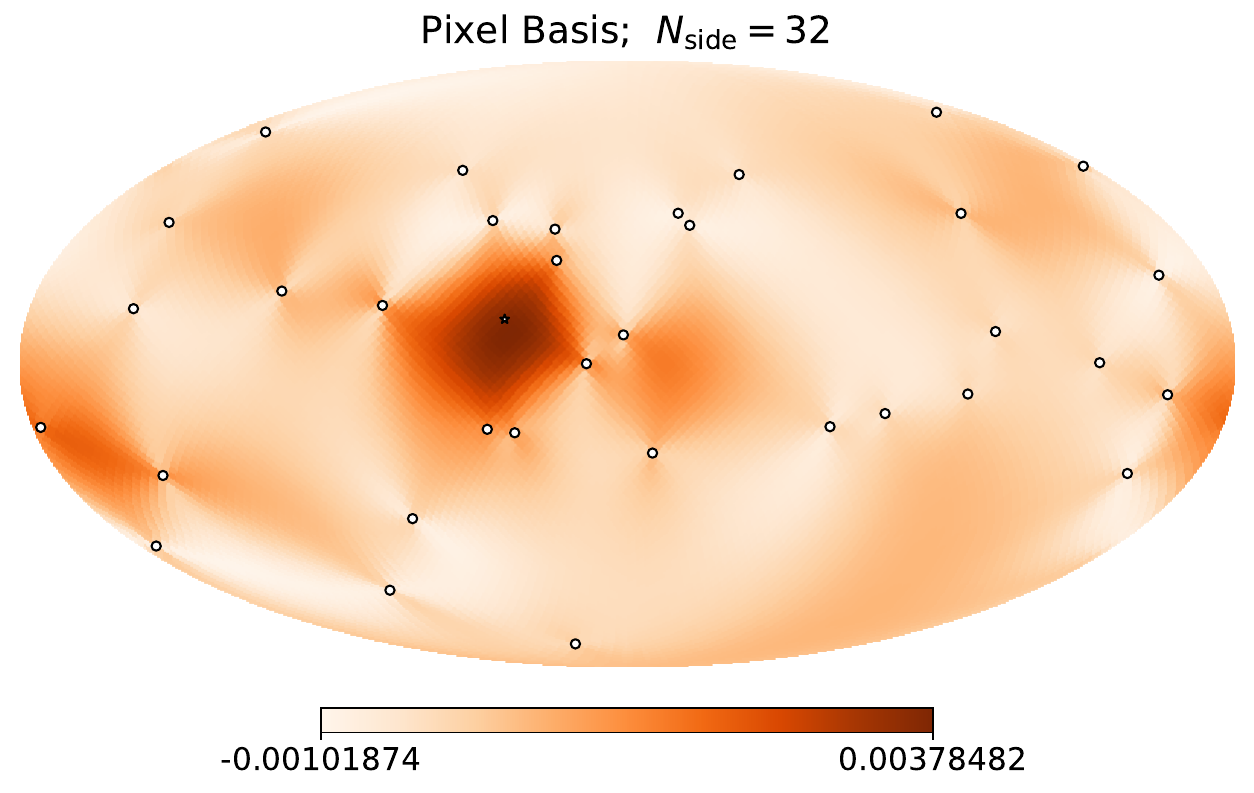}\\
    \includegraphics[scale=0.2]{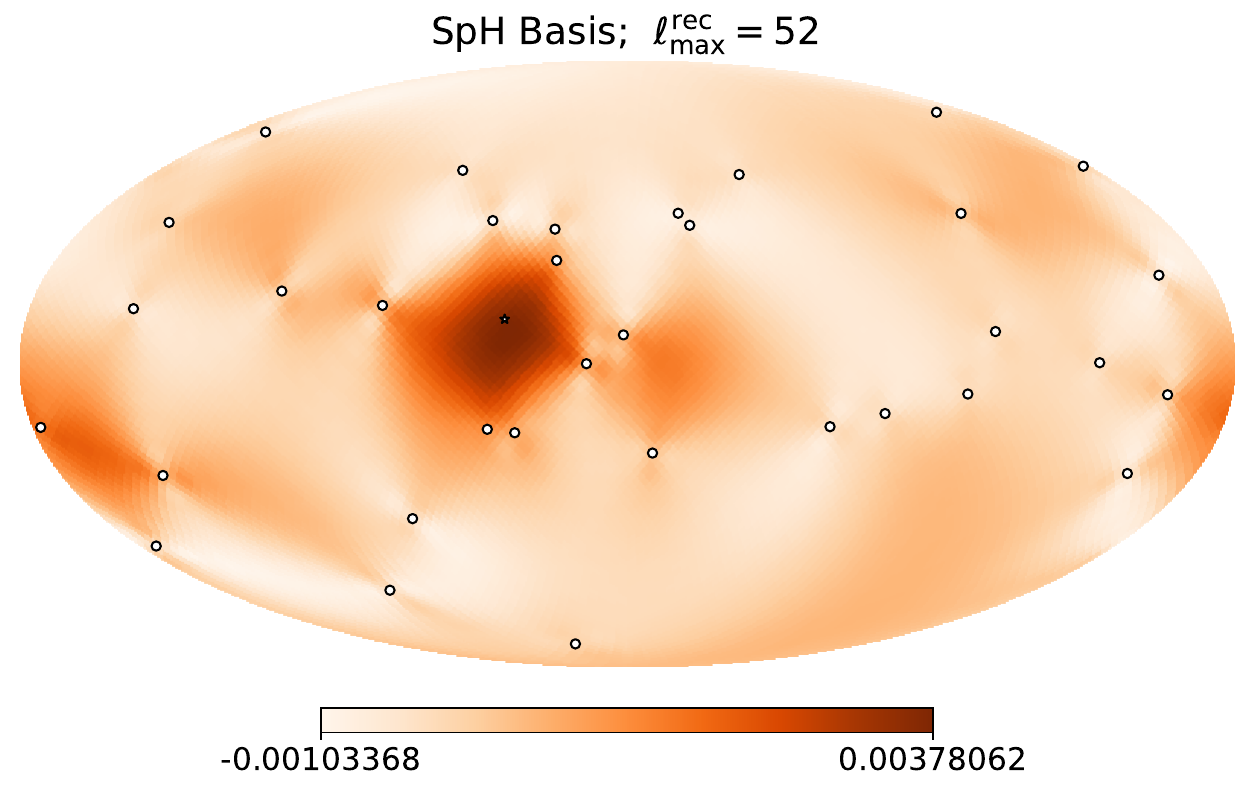}
    \includegraphics[scale=0.2]{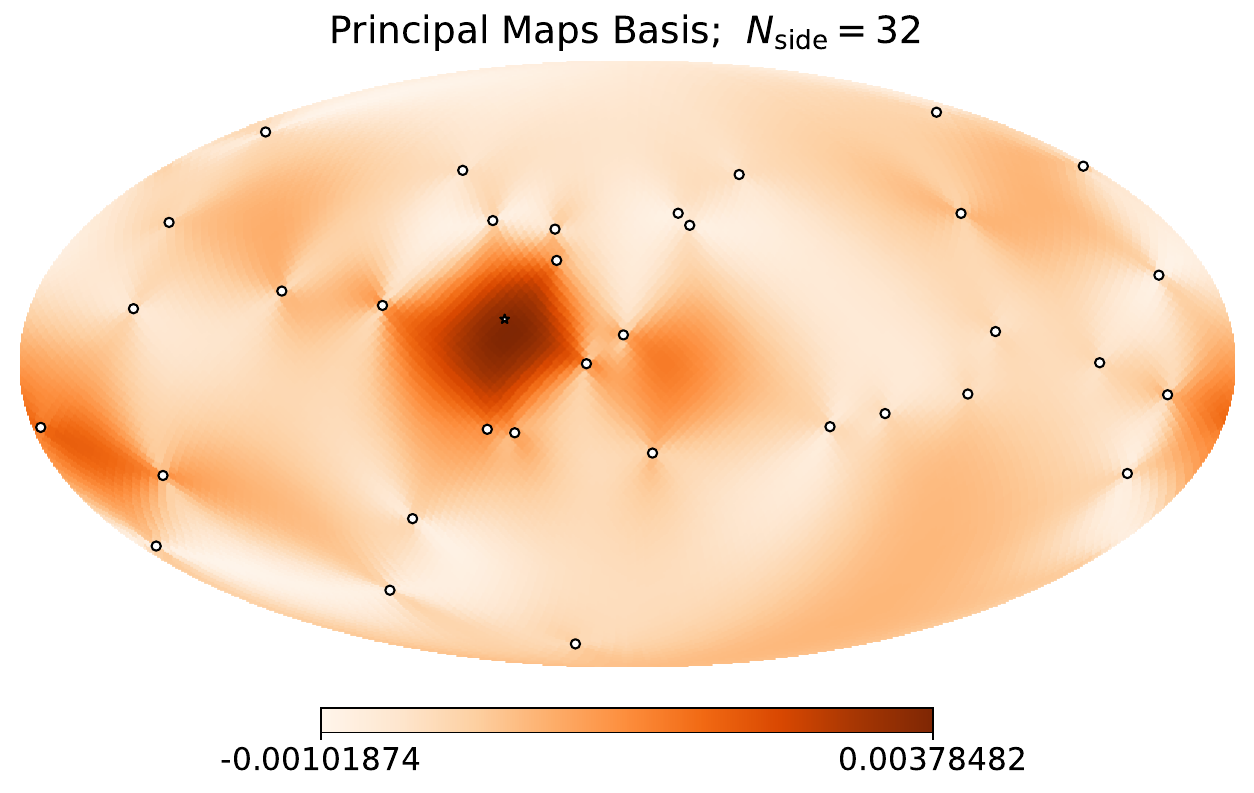}
    \caption{Reconstructed sky maps of the power $P_p$ (in units of strain$^2$) for a point source with noise ($\sigma_{ab}=1$). The top-left panel shows the dirty map in the pixel basis with $N_{\rm side}=32$. The remaining panels show reconstructions using eigenspace-based parameters with $\kappa=30$: the pixel (top right), SpH (bottom left) and principal maps (bottom right) bases. Integrating regions with $P>0.5\,P_{\rm max}$ yields a recovered total power of $\sim 1.14\pm0.21$. The choice $\kappa=30$ corresponds to retaining modes up to 33rd eigenmode, selected to limit noise amplification.
    }
    \label{fig:pointSource_Noise}
\end{figure}

In this article, we show analytically that the eigenvalues and eigenmaps from the pixel-basis and the SpH Fisher matrices are directly related to those constructed using the formalism of Refs.~\cite{Ali-Haimoud:2020ozu,alihaimoud2021} (see Sec.~\ref{s:Fisher_eigenmap} and Table~\ref{tab:cleaned_gwb}), with numerical equivalence previously demonstrated in Ref.~\cite{Agarwal:2026nxa}. Notably, the principal maps can be obtained far more efficiently using the formalism by~\citet{Ali-Haimoud:2020ozu,alihaimoud2021}, as it requires diagonalizing only an $N_{\rm pair}\times N_{\rm pair}$ matrix rather than the full high-resolution Fisher matrix [when $N_{\rm pix}\gg N_{\rm pair}$ or $(\lres+1)^2\gg N_{\rm pair}$]. We explicitly demonstrate the equivalence of map making with three bases---pixel, SpH, and principal maps---provided that the same information content is used in the reconstruction (see Sec.~\ref{s:simulation}). When the information content is mismatched, the results do not agree. We illustrate this equivalence and the origin of discrepancies for three classes of anisotropy: pointlike sources (Figs.~\ref{fig:pointSource_noNoise} and~\ref{fig:pointSource_Noise}), extended sources with deterministic anisotropy (Fig.~\ref{fig:extendedSource_noNoise}), and statistically isotropic anisotropy (Fig.~\ref{fig:statsIso_noNoise}), additionally characterized by an angular power spectrum (Fig.~\ref{fig:clRec_noNoise}). We conclude with brief remarks on the construction of principal maps for ground-based detectors (Sec.~\ref{s:Gb_det}) and a summary of our results (Sec.~\ref{s:summary}).

\section{Fisher matrix and ways to construct principal maps}
\label{s:Fisher_eigenmap}

Recently, to map the observable anisotropy, principal maps were proposed as a basis~\cite{Ali-Haimoud:2020ozu,alihaimoud2021}. These continuous principal maps, $M_n(\hatom)$, are constructed as linear combination of the PTR maps, $\gamma_{ab}(\hatom)$~\cite{Mingarelli:2013dsa,Ali-Haimoud:2020ozu,alihaimoud2021}:
   \begin{equation}
       M_n(\hatom) = \sum_{ab}  M_{n;ab}\, \frac{\gamma_{ab}(\hatom) }{\sigma_{ab}}
   \end{equation}
  (where, $\sigma_{ab}$ denote uncertainty in pulsar pair cross-correlation measurements;) with the requirement that these maps are orthogonal and span the observable space of the Fisher kernel, $F(\hatom,\hatom')$:
   \begin{equation}
\label{e:eigenMapConstruct}
   \begin{aligned}
        \int \d^2\hatom\, \d^2\hatom'\, M_n(\hatom)\, F(\hatom,\hatom')\,M_{n'}(\hatom')&=\lambda_{n}\,\delta_{nn'}\,,\\
    \sum_{ab}\sum_{cd}  M_{n;ab}\,  F_{ab,cd}  \,M_{n';cd}&=\lambda_{n}\,\delta_{nn'}\,,
   \end{aligned}
   \end{equation}  
   where $F_{ab;cd}$ (denoted by $F_{IJ}$ in Ref.~\cite{alihaimoud2021}) is given by
   \begin{equation}\label{e:Fmat_PTR}
       F_{ab;cd} \equiv \int \d^2\hatom\, \d^2\hatom'\,\frac{\gamma_{ab}(\hatom)\, \gamma_{cd}(\hatom')}{\sigma_{ab}\,\sigma_{cd}}\,F(\hatom,\hatom')\,.
   \end{equation}
   i.e., Fisher kernel projected in the pairwise timing response basis~\cite{Ali-Haimoud:2020ozu,alihaimoud2021}. Then, $\{M_{n,ab}\}$ and $\{\lambda_n\}$ represent eigenvectors and eigenvalues of a $N_{\rm pair}\times N_{\rm pair}$ matrix with elements $F_{ab;cd}$. (See Table~\ref{tab:cleaned_gwb} for details.)
   
In practice, the construction of matrix elements $F_{ab;cd}$ can be obtained only by discretizing the sky and performing a numerical integration. The pixel size is chosen to be sufficiently small to capture all significant information. A practical approach is to vary the pixel resolution and verify that the eigenvalue spectrum and eigenvectors converge, ensuring the discretization does not lose any information~\cite{Agarwal:2026nxa}. Additionally, the number of pixels does not need to match $N_{\rm pair}$ exactly; it can be much larger depending on the structure of the Fisher kernel (and the pulsar configuration comprising a PTA). 

Now, with a suitable manipulation of~\eqref{e:eigenMapConstruct} (see Appendix~\ref{s:appendixA}), we obtain
\begin{equation}\label{e:FisherEigenPix}
    \sum_q F(\hatom_p,\hatom_q) M_n (\hatom_q) = \frac{\sqrt{\lambda_n}}{\Delta\hatom} M_n (\hatom_p)\,,
\end{equation}
where $M_n(\hatom_p)$ denotes the pixelized representation of the continuous principal map $M_n(\hatom)$, and $\Delta\hatom$ denotes the solid angle subtended by each pixel of the {\tt HEALPix} grid. The eigenvalues of the pixel basis Fisher matrix are therefore $\frac{\sqrt{\lambda_n}}{\Delta\hatom}$.

Turning now to the SpH basis, the Fisher kernel can be written as
\begin{equation}
\begin{aligned}
F(\hatom,\hatom') = \sum_{\ell m \ell' m'} F_{\ell m,\ell'm'}\, Y^*_{\ell m}(\hatom)\,Y_{\ell' m'}(\hatom')\,.\\
\end{aligned}
\end{equation}
Then principal maps $M_n(\hatom)$ satisfy the eigenvalue equation~\eqref{e:FisherEigenPix},
\begin{equation}
   \begin{aligned}
   \sum_q \sum_{\ell m \ell' m'} F_{\ell m,\ell'm'}\, Y^*_{\ell m}(\hatom_p)\,Y_{\ell' m'}(\hatom_q)\, M_n (\hatom_q) \\= \frac{\sqrt{\lambda_n}}{\Delta\hatom} M_n (\hatom_p)\,.
   \end{aligned}
   \end{equation}
  Expressed in harmonic coefficient form, this becomes
   \begin{equation}
        \, \sum_{\ell m } \,M^*_{n;\ell m}\,F_{\ell m,\ell'm'}\, =\Delta\hatom\,\sqrt{\lambda_n}\,M_{n;\ell' m'}\,.
   \end{equation}
The above equation shows that the SpH basis representation of principal maps $M_{n;\ell m}=\sum_p {\Delta \hatom}\, Y^*_{\ell m}(\hatom_p) \, M_n(\hatom_p)$ are eigenvectors of SpH basis Fisher matrix with eigenvalues $\Delta\hatom\,\sqrt{\lambda_n}$.

This shows explicitly that eigenvectors of the Fisher matrix in any basis corresponds to the $N_{\rm pair}$ principal maps spanning the observable space, with eigenvalues related to $\lambda_n$ quantifying their sensitivity. The number of pixels or the number or SpH multipoles used in the analysis does not need to be equal to $N_{\rm pair}$. It can be significantly larger, depending on the structure of the Fisher matrix. What matters is that the chosen basis adequately captures all the observable information encoded in the Fisher matrix.

Note that currently $M_n(\hatom_p)$ are orthogonal but not unit norm vectors and can be easily normalized to 1. We redefine principal maps as 
\begin{equation}
    M_n(\hatom) \equiv \frac{M_n(\hatom)}{\sqrt{\int\d^2\hatom\, M^2_n(\hatom)}}\,.
\end{equation}

\section{Simulations}
\label{s:simulation}

Here, we consider a PTA configuration of 34 pulsars distributed in a statistically uniform distribution. Their sky locations are indicated by circles in Fig.~\ref{fig:pointSource_noNoise}. This is the same array configuration studied in Ref.~\cite{Agarwal:2026nxa}.

We consider two reconstruction setups for comparison: (i) In the eigenspace-converged setup: we perform the reconstruction in both the pixel and SpH bases using the parameters determined in~\cite{Agarwal:2026nxa} based on eigenspace convergence. Specifically, we adopt $N_{\rm side}=32$ and $\lrec=52$. The same choice $N_{\rm side}=32$ is also used for constructing principal maps via~\cite{Ali-Haimoud:2020ozu,alihaimoud2021} formalism. (ii) In the counting-based setup, we instead choose parameters motivated by counting argument. For 34 pulsars, the number of independent cross-correlation is $N_{\rm pair}=561$. We therefore select $N_{\rm side}=4$, for which the number of pixel satisfies $12\,N^2_{\rm side}\leq561$. Similarly, requiring $(\lrec+1)^2\leq561$ gives $\ell_{\rm max}^{\rm rec}=22$.

The normalized singular value spectra, computed numerically using the bases and corresponding parameters discussed above, are shown in Fig.~\ref{fig:SingularValue} for the pulsar configuration in Fig.~\ref{fig:pointSource_noNoise}. The spectra obtained using counting-based truncation are shown for the pixel (blue solid), SpH (blue dashed), and PTR (pink solid), while the converged spectra are shown for the pixel (orange solid), SpH (black solid), and PTR (green dashed) bases. Some of these results reproduce those presented in Fig.~3 of Ref.~\cite{Agarwal:2026nxa}. Following Ref.~\cite{Agarwal:2026nxa}, we take the spectrum obtained with the PTR basis at $N_{\rm side}=32$ as a reference. The same converged spectrum is recovered using the pixel and SpH basis with $N_{\rm side}=32$ and $\lrec=52$, respectively. In contrast counting-based truncation fails to capture some significant singular values. Since the angular values quantify the information encoded in the Fisher matrix by the PTA configuration, omitting modes associated with these singular values can lead to reconstruction artifacts if the underlying signal has non-negligible contribution along those modes.

We simulate the cross-correlations $\rho_{ab}$ by sampling from a Gaussian distribution according to the properties described in Table~\ref{tab:cleaned_gwb}. We then construct the corresponding dirty map and Fisher matrix. To obtain clean map, we employ the regularized reconstruction method described in Table~\ref{tab:cleaned_gwb} and Ref.~\cite{Agarwal:2026nxa}, for a selected condition number threshold. The results are depicted in Figs.~\ref{fig:pointSource_noNoise},~\ref{fig:extendedSource_noNoise},~\ref{fig:statsIso_noNoise},~\ref{fig:clRec_noNoise}, and~\ref{fig:pointSource_Noise}.

We note that the simulations performed in Figs. \ref{fig:pointSource_noNoise},~\ref{fig:extendedSource_noNoise},~\ref{fig:statsIso_noNoise}, and~\ref{fig:clRec_noNoise} by setting $\sigma_{ab}=0$ are well defined, as explained in Ref.~\cite{Agarwal:2026nxa}. The reason is that the clean map estimator becomes independent of $\sigma^2$ once the same noise level across pulsar pairs limit is imposed.
In this limit, the estimator variance vanishes because the Fisher information matrix diverges.

The angular power spectrum is obtained using standard estimator defined as~\cite{Agarwal:2026nxa}
    \begin{equation}
        \begin{aligned}
           \hat{C}_\ell = \frac{1}{2\ell+1} \sum_{m} \,|\hat{P}_{\ell m}|^2\,.
        \end{aligned}
    \end{equation}

We note that when eigenspace-based search parameters are used, the reconstructed anisotropy is equivalent across considered bases. The primarily computational cost arises from diagonalizing the Fisher matrix, whose dimension depends on the chosen basis: for the pixel, SpH and principal maps basis, it scales as, i.e., $N_{\rm pix}=12{,}288$, $(\lrec+1)^2=2809$ and $N_{\rm pair}=561$. Consequently, the principal map approach is the most computationally efficient.

\section{Comment on mapping anisotropy with ground-based detectors}
\label{s:Gb_det}

We note that this study of basis equivalence is relevant for mapping GWB anisotropy in the audio-frequency band probed by ground-based interferometers. In that context, anisotropy is mainly decomposed in the pixel~\cite{Mitra:2007mc} and spherical harmonic bases~\cite{Thrane:2009fp}. A similar analysis could be performed by replacing the PTR $\gamma$ and corresponding Fisher matrices with those appropriate for ground-based detectors. Although no study has yet explored a reconstruction of principal maps using a matrix analogous to $N_{\rm pair}\times N_{\rm pair}$ matrix for ground-based interferometers, eigenmaps indirectly play a role when truncated SVD regularization is applied~\cite{Thrane:2009fp,Agarwal:2021gvz,Floden:2022scq}. 

An analogy can be drawn from Sec.~\ref{s:Fisher_eigenmap}. Unlike PTAs, the baseline response of ground-based interferometers to the GWB, characterized by overlap reduction function (ORF), depends upon signal frequency $f$, sidereal time $t_s$, and detector baseline $I$~\cite{Mitra:2007mc,Thrane:2009fp,Thrane:2015aua,Ain:2015lea}. In this case, the principal maps can be constructed as
   \begin{equation}
       M_n(\hatom) = \sum_{Ift_s}  M_{n;Ift_s}\, \frac{\gamma_{Ift_s}(\hatom)}{\sigma_{Ift_s}}
   \end{equation}
   with the requirement that these maps are orthogonal and span the observable space of Fisher kernel:
   \begin{equation}
\label{e:eigenMapConstruct_LVK}
   \begin{aligned}
        \int \d^2\hatom\, \d^2\hatom'\, M_n(\hatom)\, F(\hatom,\hatom')\,M^*_{n'}(\hatom')&=\lambda_{n}\,\delta_{nn'}\,,\\
    \sum_{II'}\sum_{ff'}\sum_{t_st_s'}  M_{n;Ift_s}\,  F_{Ift_s,I'f't'_s}  \,M^*_{n';I'f't'_s}&=\lambda_{n}\,\delta_{nn'}\,,
   \end{aligned}
   \end{equation}  
   where Fisher kernel is given by~\cite{Ain:2015lea}
   \begin{equation}
       F(\hatom,\hatom') = \sum_{Ift_s} \frac{\gamma^*_{Ift_s}(\hat{\Omega}) \, \gamma_{Ift_s}(\hat{\Omega}')}{\sigma^2_{Ift_s}}
   \end{equation}
   and $F_{Ift_s,I'f't'_s}$ is given by
   \begin{equation}
       F_{Ift_s,I'f't'_s} \equiv \int \d^2\hatom\, \d^2\hatom'\,\frac{\gamma_{Ift_s}(\hatom)}{\sigma_{Ift_s}}\, \frac{\gamma^*_{I'f't'_s}(\hatom')}{\sigma_{I'f't'_s}}\,F(\hatom,\hatom')\,.
   \end{equation}
   (Additionally, we will have to account for negative frequency contribution carefully.) Here, the size of matrix $F_{Ift_s,I'f't'_s}$ is much larger than the Fisher matrix in pixel and SpH basis. This is because there are typically $\sim 5\times10^4$ frequency bins and $\sim 10^3$ time sidereal segments to analyze for a broadband directional search, while the number of baselines is comparatively small (e.g., three baselines for the current LIGO-Hanford and LIGO-Livingston and Virgo network). Consequently, unlike the PTA case, obtaining principal maps by diagonalizing the Fisher matrix in the pixel ($N_{\rm side}=16$;~\cite{Mitra:2007mc}) and SpH ($\lrec=15$; \cite{Thrane:2009fp,Floden:2022scq}) bases is computationally much faster. However, using the principal map basis for Bayesian parameter estimation of an anisotropic background~\cite{Tsukada:2022nsu} [where direct inference over the full set of $P_{\ell m}$ or $P(\hatom)$ coefficients becomes challenging due to the large parameter space] will be an interesting direction to explore.
   
   On the other hand, the narrowband directional search---such as all-sky-all-frequency analyses~\cite{KAGRA:2021rmt,Agarwal:2023lzz}---may benefit from constructing principal maps using the Fisher matrix in the ORF basis. In this case, given the angular resolution of the search increases with frequency, potentially requiring a larger number of pixels to decompose the sky than the number of available sidereal segments for a given detector baseline. Exploring this regime and assessing the computational advantages of the principal map construction remains another interesting direction for future work. 

\section{Conclusions}
\label{s:summary}

In this work, we have examined the relationship between different approaches to mapping anisotropies in the GWB, focusing on pixel, SpH and principal map formulations. We have shown that these methods are fundamentally equivalent, provided they retain the same information content, i.e., span the same eigenspace of the Fisher matrix described by the PTA response and noise properties.

Our results clarify that the apparent difference between reconstruction methods do not arise from the choice of basis itself, but from how the underlying parameter space is truncated. In particular, truncation schemes based solely on counting arguments may fail to capture the full set of informative modes leading to incomplete or based reconstructions. By contrast, constructions that ensure the convergence of the Fisher matrix eigenspectrum recover consistent and basis-independent description of observable anisotropy.

These findings have practical implications for future PTA anisotropy searches. Extension of this framework to realistic PTA datasets, as well as to ground-based interferometers, will be important directions for future work. In addition, beyond truncated SVD regularization, alternative regularization strategies informed by the expected source anisotropy may be considered in future work.

\section{DATA AVAILABILITY}
There are no publicly available research data or software supporting this manuscript. Requests for further
information or data should be sent to the authors.

\begin{acknowledgements}
The author gratefully acknowledges Joe Romano for bringing the counting argument problem to attention, for insightful discussion on the role of basis, and for valuable comments on earlier drafts of this article. This work greatly benefitted from discussion with Yacine Ali-Ha\"imoud and Tristan L. Smith regarding the construction of principal maps formalism discussed in Refs.~\cite{Ali-Haimoud:2020ozu,alihaimoud2021}. The author is grateful to Chiara Mingarelli for detailed comments on the manuscript, which significantly improved its clarity and readability. D.A.~acknowledges financial support from 
NSF Physics Frontier Center Grant No. PFC-2020265 and 
start-up funds from the University of Texas Rio Grande Valley.
Some of the results in this article have been derived using the {\tt HEALPY} and {\tt HEALPix} packages.
\end{acknowledgements}

\appendix

\section{}
\label{s:appendixA}

\subsection{Relation between eigenspace of Fisher matrix in PTR and pixel bases}

Let us start with the discrete representation of Fisher kernel in PTR basis~\eqref{e:eigenMapConstruct} given as
\begin{equation}
    \bm{F}^{\rm PTR} \approx (\bm{\Gamma} \bm{\Gamma}^T) (\bm{\Gamma}\bm{\Gamma}^T) (\Delta\hatom)^2\,,
\end{equation}
where we have defined matrix $\bm{\Gamma}\equiv \frac{\gamma_{ab,p}}{\sigma_{ab}}$. The Fisher matrix in pixel basis can be written as (see Table~\ref{tab:cleaned_gwb})
\be
\bm{F}^{pix}=\bm{\Gamma}^T \bm{\Gamma}\,.
\ee
 Then, $n$-th eigenvalue $(\lambda_n)$ and its corresponding eigenvector $(\bm{M}_n)$ for matrix $\bm{F}^{\rm PTR}$ satisfy~\eqref{e:eigenMapConstruct}
    \begin{equation}
    \begin{aligned}
    \bm{F}^{\rm PTR} \bm{M}_n &= \lambda_n\, \bm{M}_n\\
        (\bm{\Gamma} \bm{\Gamma}^T) (\bm{\Gamma}\bm{\Gamma}^T) (\Delta\hatom)^2\, \bm{M}_n&= \lambda_n\, \bm{M}_n\\
        (\bm{\Gamma} \bm{\Gamma}^T) \bm{M}_n &= \frac{\sqrt{\lambda_n}}{\Delta\hatom} \bm{M}_n\\
    ( \bm{\Gamma}^T   \bm{\Gamma}) (\bm{\Gamma}^T \bm{M}_n) &= \frac{\sqrt{\lambda_n}}{\Delta\hatom} (\bm{\Gamma}^T \bm{M}_n)\,.
    \end{aligned}
\end{equation}
Then, $\bm{F}^{pix} \bm{M}_n = \frac{\sqrt{\lambda_n}}{\Delta\hatom} \bm{M}_n$ where $\bm{M}_n=\bm{\Gamma}^T \bm{v}_n$.

\subsection{Alternative way to characterize eigenspace of Fisher matrix in PTR basis}

Here we show that same principal maps can also be constructed using a matrix defined as
\begin{equation}
    G_{ab,cd} \equiv \int \d^2 \hatom\,\frac{\gamma_{ab}(\hatom)\gamma_{cd}(\hatom)}{\sigma_{ab}\sigma_{cd}}\,,
\end{equation}
which is related to the matrix $\bm{F}$~\eqref{e:Fmat_PTR}
\be
F_{ab,cd} \equiv \int \d^2 \hatom\, \d^2 \hatom'\, F(\hatom,\hatom')\,\frac{\gamma_{ab}(\hatom)\,\gamma_{cd}(\hatom')}{\sigma_{ab}\sigma_{cd}}\,,
\ee
as (using Fisher kernel from Table~\ref{tab:cleaned_gwb})
\be
\ba
F_{ab,cd} &= \int \d^2 \hatom\, \d^2 \hatom'\,\sum_{ef} \,\frac{\gamma_{ef}(\hatom)\gamma_{ef}(\hatom')}{\sigma^2_{ef}}\,\frac{\gamma_{ab}(\hatom)\, \gamma_{cd}(\hatom')}{\sigma_{ab}\,\sigma_{cd}} \\
&= \sum_{ef} G_{ab,ef} G_{ef,cd}\\
\bm{F} &= \bm{G}\bm{G}^T\,.
\ea
\ee
If we decompose $\bm{G}$ matrix in terms of its eigenvalues and eigenvectors
\be
\ba
\sum_{ef} G_{ab,ef}M^{ef}_{n} = \Sigma_n\,M^{ab}_{n}\\
\bm{G} \bm{M}= \bm{M} \bm{\Sigma}\,,
\ea
\ee
then,
\be
\ba
\bm{F} &= \bm{M} \bm{\Sigma}\bm{M}^T\bm{M} \bm{\Sigma}\bm{M}^T\\
 &= \bm{M} \bm{\Sigma}^2\bm{M}^T\,.
\ea
\ee
Hence, the eigenvalues of matrix $\bm{F}$ are the square of those of matrix $\bm{G}$ and the eigenvectors are the same.

\bibliography{ref}

\end{document}